\documentclass[12pt]{article}
\usepackage{bm}
\usepackage{amsmath}
        \numberwithin{equation}{section}
\usepackage{amssymb}
\usepackage[dvips]{graphicx}
\usepackage{bbold}
\usepackage{multirow}
\usepackage{enumerate}
\usepackage{ulem}
\usepackage{color}

\definecolor{refkey}{rgb}{1,0,0}
\definecolor{labelkey}{rgb}{0,0,1}

\usepackage[utf8]{inputenc}
\begin{document}

\begin{titlepage}
\begin{flushright}
TIT/HEP-665 \\
April,  2018
\end{flushright}
\vspace{0.5cm}
\begin{center}
{\Large \bf
Quantum periods for 
$\mathcal{N}=2$ $SU(2)$ SQCD 
around the superconformal point
}
\lineskip .75em
\vskip 2.5cm
{\large  Katsushi Ito and Takafumi Okubo }
\vskip 2.5em
 {\normalsize\it Department of Physics,\\
Tokyo Institute of Technology\\
Tokyo, 152-8551, Japan}
\vskip 3.0em
\end{center}
\begin{abstract}
We study the Argyres-Douglas theories realized at the superconformal point in the Coulomb moduli space of  $\mathcal{N}=2$  supersymmetric $SU(2)$ QCD with $N_f=1,2,3$ hypermultiplets in the Nekrasov-Shatashvili limit of the Omega-background.
The Seiberg-Witten curve of the theory is quantized in this limit and  the periods
receive the quantum corrections.
By applying the WKB method for the quantum Seiberg-Witten curve, we
calculate the quantum corrections to the Seiberg-Witten periods around the superconformal point up to the fourth order in the parameter
of the Omega background.
\end{abstract}
\end{titlepage}
\baselineskip=0.7cm

\section{Introduction}
A large class of ${\cal N}=2$ supersymmetric gauge theories has a superconformal fixed point at strong coupling in the Coulomb moduli space, where mutually non-local BPS states become massless. 
This theory becomes an interacting ${\cal N}=2$ superconformal field theory, which is called the Argyres-Douglas (AD) theory \cite{Argyres:1995jj,Argyres:1995xn}. 
The BPS spectrum of the AD theory can be studied by the Seiberg-Witten (SW) curve,
which are obtained from degeneration of the curve of ${\cal N}=2$ gauge theories \cite{Argyres:1995jj,Argyres:1995xn,Eguchi:1996vu}. 
The dynamics of AD theories
is an interesting subject of recent studies from the viewpoint of M5-branes compactified on a punctured Riemann surface \cite{Gaiotto:2009we,Gaiotto:2009hg,Xie:2012hs} and its relation to two-dimensional conformal field theories \cite{Beem:2013sza,Liendo:2015ofa,Cordova:2015nma,Buican:2015ina}. 

 In the weak coupling region, one can compute the partition function of ${\cal N}=2$ gauge theories based on the microscopic Lagrangian in the $\Omega$-background, which deforms four-dimensional spacetime by the torus action with two parameters $(\epsilon_1,\epsilon_2)$ \cite{Nekrasov:2002qd,Nekrasov:2003rj}. 
The partition function are related to conformal blocks of two-dimensional conformal field theories \cite{Alday:2009aq,Gaiotto:2009ma}, the partition functions of 
topological strings \cite{Huang:2009md,Huang:2012kn}, and the solutions of the Painl\'eve equations \cite{Bonelli:2016qwg}, where the $\Omega$-deformation parameters 
enter into the formulas of  the central charges and the string coupling. 
It would be interesting to study the effects of the $\Omega$-deformations in the strong coupling region.
However in the strong coupling region such as the superconformal point, we have no appropriate microscopic Lagrangian.
In the case of the self dual $\Omega $-background with $\epsilon _1=-\epsilon _2$, the Argyres-Douglas theories have been studied by using the holomorphic anomaly equation \cite{Huang:2009md,Huang:2011qx} and the E-strings \cite{Sakai:2016jdi}.

The purpose of this paper is to study the Argyres-Douglas theories in the $\Omega$-background realized at the superconformal point of ${\cal N}=2$ supersymmetric gauge theories. In particular, we consider the Nekrasov-Shatashvili (NS) limit \cite{Nekrasov:2009rc} of the $\Omega $ background where one of the deformation parameters $\epsilon_2$  is set to be zero. 
In this limit the SW curve becomes a differential equation which is obtained by the canonical quantization procedure of the symplectic structure induced by the SW differential. 
The Planck constant $\hbar$ corresponds to the remaining deformation parameter $\epsilon_1$.
The WKB solution of the differential equation gives the $\Omega$-deformation of the SW periods which is the main subject of this paper.

The quantum SW curve has been studied for ${\cal N}=2$ theories in the weak coupling regions. 
A simple example is $SU(2)$ pure Yang-Mills theory where the quantum SW curve becomes the Schr\"odinger equation with the sine-Gordon potential \cite{Mironov:2009uv} and the WKB solution is shown to agree with that obtained from the NS limit of the Nekrasov function.
The expansion of the periods around the massless monopole point in the Coulomb moduli space has been studied in \cite{He:2010xa}.
For ${\cal N}=2$ $SU(2)$ SQCD with $N_f\leq 4$ hypermultiplets, the WKB solutions of the quantum SW curves have been studied in \cite{Zenkevich:2011zx} in the weak coupling region, while in the strong coupling region  the solutions around the massless monopole point  have been studied in \cite{Ito:2017iba}.
Generalization to other ${\cal N}=2$ theories and their relations to the Nekrasov partition functions have been studied extensively \cite{Mironov:2009dv,Zenkevich:2011zx,Popolitov:2010bz,Beccaria:2016wop,He:2016khf}.

In this paper we will study the quantum SW periods around the superconformal point of the moduli space of ${\cal N}=2$ $SU(2)$ SQCD with $N_f=1,2,3$ hypermultiplets. The SW curve degenerates into a simpler curve which represents the SW curve of the Argyres-Douglas theory. 
We will calculate the WKB solution of the quantum SW curve of the AD theory and compute the quantum corrections up to the fourth order in $\hbar$. 

This paper is organized as follows:
In Section 2, we review the SW curve and the SW differential near the superconformal point of the $\mathcal{N}=2$ $SU(2)$ SQCD.
In Section 3, we quantize  the SW curve of the AD theories and derive the differential equations satisfied by quantum periods.
In Section 4, we calculate the quantum corrections to the SW periods near the  superconformal point, which are
expressed in terms of
the hypergeometric function.
Section 5 is devoted to conclusions and discussion. 
In the Appendix, we present detailed analysis of the fourth order terms in the quantum SW periods for the $N_f=3$ AD theory.

\section{Seiberg-Witten curve at the superconformal point}
In this section we study the Argyres-Douglas theory which appears at the superconformal point in the moduli
space of ${\cal N}=2$ $SU(2)$ SQCD with $N_f=1,2,3$ hypermultiplets.
We begin with the Seiberg-Witten curve for the $\mathcal{N}=2$ $SU(2)$ gauge theory with $N_f(=1,2,3)$ hypermultiplets which is given by
\begin{align} \label{eq:swc}
C(p) -\frac{\Lambda _{N_f}^{2-\frac{N_f}{2}}}{2}\left( z+\frac{G(p)}{z} \right)=0,
\end{align}
where $\Lambda _{N_f}$ is the QCD scale parameter. $C(p)$ and $G(p)$ are defined by
\begin{align}
C(p)&=\left\{ \begin{array}{cc}
p^2-u, & N_f=1,\\
p^2-u+\frac{\Lambda _2^2}{8}, & N_f=2, \\
p^2-u +\frac{\Lambda _3}{4} \left( p+\frac{m_1+m_2+m_3}{2}\right), & N_f=3,
\end{array}
\right.\\
\label{eq:hyper}
G(p)&=\prod _{i=1}^{N_f} (p+m_i),
\end{align}
where $u$ is the Coulomb moduli parameter and $m_1, \ldots ,m_{N_f}$ are the mass parameters of the hypermultiplets.
The SW differential is defined by
\begin{align} \label{swdiff}
\lambda _{\text{SW}}= p \left( d \log G(p) -2 d \log z\right) .
\end{align}
The SW periods $\Pi ^{(0)}:=(a^{(0)}, a_D^{(0)})$ are
\begin{align}
a^{(0)}(u) =\oint _\alpha \lambda _{\text{SW}} , \qquad \qquad a_D^{(0)}(u) =\oint _\beta \lambda _{\text{SW}}
\end{align}
where $\alpha $ and $\beta $ are the canonical one-cycles on the curve. 
Here the superscript $(0)$ refers the ``undeformed" (or classical) period.
The SW curve (\ref{eq:swc}) can be written into the standard form \cite{Hanany:1995na}
\begin{align} \label{eq:swc2}
y^2 &=C(p)^2 -\Lambda _{N_f}^{4-N_f} G(p)
\end{align}
by introducing 
\begin{align} \label{transf:swc}
y=\Lambda _{N_f} ^{2-\frac{N_f}{2}} z-C(p) .
\end{align}
The SW differential (\ref{swdiff}) is expressed as
\begin{align}
\lambda_{SW}&=pd\log\left({C(p)-y\over C(p)+y}\right).
\end{align}
The $u$-derivative of the SW differential becomes the holomorphic differential: 
\begin{align} \label{swdiff_du}
\frac{\partial \lambda _{\text{SW}}}{\partial u} =\frac{2 \partial _u z}{z} dp +d(*)={2dp\over y}+d(*)
\end{align}
where $\partial _u:=\frac{\partial }{\partial u}$.
Differentiating the SW period $\Pi^{(0)}$ with respect to $u$, one obtains the periods for the curve:
\begin{align}
\partial _u a^{(0)}(u) =\oint _\alpha \frac{2 \partial _u z}{z} dp =\oint _\alpha \frac{2 }{y}dp, \qquad \qquad \partial _u a_D^{(0)}(u) =\oint _\beta \frac{2 \partial _u z}{z} dp =\oint _\beta \frac{2 }{y}dp.
\end{align}
The period $\partial_u\Pi ^{(0)}$ is evaluated as the elliptic integral. 
For the curve of the form $y^2=\prod_{i=1}^4(x-e_i)$, it is convenient to introduce the variables
\begin{align}\label{eq:dd}
D=&\sum_{i<j}e_i^2 e_j^2-6\prod_{i=1}^4 e_i-\sum_{i<j<k}(e_i^2e_j e_k+e_i e_j^2 e_k+e_i e_j e_k^2), \\
w=&-\frac{27 \Delta }{4 D^3},
\end{align}
where $\Delta $ is of the discriminant
\begin{align} \label{discriminant:swc}
\Delta &=\prod _{i <j}(e_i-e_j)^2 .
\end{align}
and $w$ is inverse of the modular $J$-function of the curve \cite{Brandhuber:1996ng}.
Then it is shown that the integral $F=(-D)^{\frac{1}{4}}\int {dx\over y}$ obeys the
hypergeometric differential equation
\begin{align}
w(1-w){d^2 F\over dw^2}+(\gamma-(\alpha+\beta+1)w){dF\over dw}-\alpha\beta F=0
\label{eq:hyper1}
\end{align}
with $\alpha=\frac{1}{12}$, $\beta=\frac{5}{12}$ and $\gamma=1$.
For the SW curve (\ref{eq:swc2}) this leads to the Picard-Fuchs equation for $\Pi^{(0)}$ \cite{Ito:1995ga,Ohta:1996hq,Ohta:1996fr,Ito:2017iba}
as the third order differential equation with respect to $u$.

There are singularities on the $u$-plane where some BPS particles become massless and the discriminant $\Delta $ (\ref{discriminant:swc}) becomes zero.
We consider the superconformal or Argyres-Douglas (AD) point on the $u$-plane where mutually nonlocal BPS particles become massless \cite{Argyres:1995jj,Argyres:1995xn}.
For the $SU(2)$ theory with $N_f$ hypermultiplets, the squark and  monopole/dyon are both massless at
the AD point, where the SW curve degenerates and has higher order zero.
For the $SU(2)$ theories with $N_f=1,2,3$ hypermultiplets,
the AD points are given as follows:
For $N_f=1$, the Coulomb moduli and the mass are chosen as
\begin{align} \label{ADpoint:nf1}
u =\frac{3}{4} \Lambda _1^2, \qquad m_1= \frac{3}{4} \Lambda _1 .
\end{align}
The SW curve (\ref{eq:swc2}) becomes
\begin{align}\label{eq:swnf1}
y^2 =\left( p -\frac{2}{3} \Lambda _1 \right) \left( p +\frac{1}{2} \Lambda _1 \right) ^3.
\end{align}
For $N_f=2$, we have 
\begin{align} \label{ADpoint:nf2}
u=\frac{3}{8} \Lambda _2^2, \qquad m_1=m_2=\frac{\Lambda _2}{2} ,
\end{align}
so that the SW curve (\ref{eq:swc2}) becomes
\begin{align}
y^2=\left( p-\frac{3}{2} \Lambda _2\right) \left( p +\frac{\Lambda _2}{2} \right)^3.
\end{align}
For $N_f=3$, the superconformal point is given by
\begin{align} \label{ADpoint:nf3}
u=\frac{1}{32}\Lambda _3^2, \qquad m_1=m_2=m_3 = \frac{\Lambda _3}{8} ,
\end{align}
where the SW curve (\ref{eq:swc2}) becomes
\begin{align}
y^2= \left( p-\frac{7}{8} \Lambda _3\right)\left( p+\frac{\Lambda _3}{8} \right)^3.
\end{align}

Let us study the SW curve and the SW differential around the superconformal point.
By taking the scaling limit, we identify the operators and couplings which deform the superconformal point. Their scaling dimensions are determined by the SW curve and the fact that the SW differential has the scaling dimension one.
We first consider in the $N_f=1$ theory.
The branch point $p=- \frac{\Lambda _1}{2}$ of the curve (\ref{eq:swnf1}) corresponds to $z=\pm \frac{\Lambda _1^{\frac{1}{2}}}{2}$.
We expand the curve (\ref{eq:swc}) around $z=-\frac{\Lambda _1^{\frac{1}{2}}}{2}$ by introducing
\begin{align} 
p=&\epsilon \tilde{p}-\frac{\Lambda _1}{2}, \qquad z= \frac{i2^{\frac{1}{2}} \epsilon ^{\frac{3}{2}} }{\Lambda _1} \tilde{z} -\frac{\epsilon ^2\tilde{M}}{\Lambda _1^{\frac{1}{2}}}-\frac{\epsilon \tilde{p}}{\Lambda _1^{\frac{1}{2}}}-\frac{\Lambda _1^{\frac{1}{2}}}{2},\nonumber \\
u=&\epsilon ^3\tilde{u}+\epsilon ^2\tilde{M} \Lambda _1+\frac{3}{4}\Lambda _1^2,\qquad m_1=\epsilon ^2\tilde{M}+\frac{3}{4}\Lambda _1, \label{rescaling:nf1}
\end{align}
and consider the scaling limit $\epsilon\rightarrow 0$ with fixed $\tilde{u}$ and $\tilde{M}$.
At the leading order in $\epsilon$ we obtain the curve for the AD theory of $(A_1,A_2)$-type: 
\begin{align}\label{eq:swcadnf1}
\tilde{z}^2&=\tilde{p}^3-\tilde{M}\Lambda_1 \tilde{p}-{\Lambda_1\over2}\tilde{u}.
\end{align}

Substituting  (\ref{rescaling:nf1}) into the SW differential (\ref{swdiff}) and expanding around $\epsilon =0$, the SW differential becomes 
\begin{align} 
\lambda_{SW}&={i\epsilon^{\frac{5}{2}} \over 2^{\frac{1}{2}} \Lambda_1^{\frac{1}{2}}}\tilde{\lambda}_{SW}+\ldots, \\
\tilde{\lambda}_{SW}&:=-\frac{8}{\Lambda _1} \tilde{z}d\tilde{p}. \label{addiff:nf1}
\end{align}
We  read off the scaling dimension of $\tilde{u}$ and $\tilde{M}$ as $\frac{6}{5}$ and $\frac{4}{5}$, respectively,
from the curve (\ref{eq:swcadnf1}).
Here $\tilde{u}$ is the operator and $\tilde{M}$ is the corresponding coupling parameter.

For $N_f=2$,
defining  the new variables as 
\begin{align}
p&= \epsilon \tilde{p} -\frac{\epsilon \tilde{M}}{3} -\frac{\Lambda _2}{2}, \qquad z=\frac{i 2^{\frac{1}{2}} \epsilon ^{\frac{3}{2}}}{\Lambda _2^{\frac{1}{2}} } \tilde{z} - \epsilon \tilde{p} -\frac{2\epsilon \tilde{M} }{3}, \nonumber \\
u&= \epsilon ^2 \tilde{u} -\frac{(\epsilon \tilde{M} )^2}{3}+\Lambda  _2 \epsilon \tilde{M} +\frac{3 \Lambda _2^2}{8} , \nonumber \\
m_1&= \frac{\Lambda _2}{2}+\epsilon \tilde{M} +\epsilon ^{\frac{3}{2}} \tilde{a} ,\qquad m_2=\frac{\Lambda _2}{2}+\epsilon \tilde{M} -\epsilon ^{\frac{3}{2}} \tilde{a},
\label{rescaling:nf2}
\end{align}
and 
expanding the curve around $\epsilon =0$, we find that the curve 
 (\ref{eq:swc})
become
\begin{align} \label{adc:nf2}
\tilde{z}^2&=\tilde{p}^3-\tilde{u}\tilde{p}-{2\over3}\tilde{M}\tilde{u}
+{8\over27}\tilde{M}^3-{\tilde{C}_2\Lambda_2\over 4}.
\end{align}
Here $\tilde{u}$ is the operator, $\tilde{M}$ is the coupling and $\tilde{C}_2:=2 \tilde{a}^2$ is the Casimir invariant of the $U(2)$ flavor symmetry.
The corresponding AD theory is of $(A_1,A_3)$-type.

Substituting (\ref{rescaling:nf2}) into (\ref{swdiff}),  the SW differential around the superconformal point is
\begin{align}  \label{addiff:nf2}
\lambda _{\text{SW}} = \frac{i \epsilon ^{\frac{3}{2}}}{2^{\frac{1}{2}} \Lambda _2^{\frac{1}{2}} } \tilde{\lambda }_{\text{SW}}  +\cdots
\end{align}
up to the total derivatives where
\begin{align}\label{eq:swdnf2}
\tilde{\lambda }_{\text{SW}} = -4\tilde{z}\, d\log \left( \tilde{p}+ \frac{2}{3}\tilde{M} \right).
\end{align}
The scaling dimension of $\tilde{u}$, $\tilde{M}$ and $\tilde{C}_2$ are $\frac{4}{3}$, $\frac{2}{3}$ and $2$, respectively.

For $N_f=3$,
we define the scaling variables as
\begin{align} 
&p=\epsilon ^2 \tilde{p}-\epsilon \tilde{M}+\frac{4 \left((\epsilon \tilde{M}) ^2+\epsilon ^3 \tilde{u}\right)}{3 \Lambda _3}+\frac{16 (\epsilon \tilde{M})^3}{9 \Lambda _3^2} -\frac{\Lambda _3}{8}  ,\nonumber 
\\
&z= \epsilon ^3 i \tilde{z}- \frac{4 (\epsilon \tilde{M})^3}{3 \Lambda _3^{\frac{3}{2}} } -\frac{2 (\epsilon \tilde{M})  (\epsilon ^2 \tilde{p}) }{\Lambda _3^{\frac{1}{2}}} -\frac{\epsilon ^3 \tilde{u}}{\Lambda _3^{\frac{1}{2}}},  \nonumber 
\\
&u=\epsilon ^3 \tilde{u}-\frac{4 (\epsilon \tilde{M}) ^3}{3 \Lambda _3}+(\epsilon \tilde{M})^2+\frac{3 \Lambda  _3 \epsilon \tilde{M}}{8} +\frac{\Lambda _3^2}{32} , \nonumber \\
&m_1 =\frac{\Lambda _3}{8} +\epsilon \tilde{M} +\epsilon ^2 \tilde{c}_1, \qquad 
m_2= \frac{\Lambda _3}{8} +\epsilon \tilde{M} +\epsilon ^2 \tilde{c}_2, \qquad
m_3= \frac{\Lambda _3}{8} +\epsilon \tilde{M} -\epsilon ^2 (\tilde{c}_1+\tilde{c}_2).
\label{rescaling:nf3} 
\end{align}
and then consider the limit $\epsilon\rightarrow 0$ limit with keeping $\tilde{u}$, $\tilde{M}$, $\tilde{c}_1$ and $\tilde{c}_2$ finite.
Rescaling the curve (\ref{eq:swc})
we obtain the curve of the  AD theory of $(A_1,D_4)$ type:
\begin{align}
\tilde{z}^2=\tilde{p}^3-\tilde{p} \left(\frac{\tilde{C_2}}{2}+\frac{4 \tilde{M} \tilde{u}}{\Lambda _3}\right) -\frac{\tilde{u}^2}{\Lambda _3}-\frac{8 \tilde{M}^3 \tilde{u}}{3 \Lambda _3^2} +\frac{16 \tilde{M}^6}{27 \Lambda _3^3} -\frac{2 \tilde{C_2} \tilde{M}^2}{3 \Lambda _3}+\frac{\tilde{C_3}}{3} \label{adc:nf3}
\end{align}
where
\begin{align}
\tilde{C_2} :=&2\left( \tilde{c}_1^2+\tilde{c}_1\tilde{c}_2+\tilde{c}_2^2 \right) ,\qquad \tilde{C_3}:=-3\left(  \tilde{c}_1^2 \tilde{c}_2 +\tilde{c}_1 \tilde{c}_2^2 \right) .
\end{align}
Here $\tilde{u}$ is the operator and 
$\tilde{M}$ is the coupling.
$\tilde{C}_2$ and $\tilde{C}_3$ are the  Casimir invariants associated with the $U(3)$ flavor symmetry.
Then the SW differential (\ref{swdiff}) at the superconformal point becomes
\begin{align} \label{addiff:nf3}
\lambda _{\text{SW}}=\frac{i \epsilon ^2}{\Lambda _3^{\frac{1}{2}}} \tilde{\lambda }_{\text{SW}}  +\cdots
\end{align}
up to the total derivatives where
\begin{align}\label{eq:swadnf3}
\tilde{\lambda }_{\text{SW}} =i\Lambda_3^{\frac{1}{2}}\left\{ 2 \tilde{p}\, d\log \left(i \tilde{z}-\frac{2 \tilde{M} \tilde{p}}{\Lambda _3^{\frac{1}{2}}}-\frac{4 \tilde{M}^3}{3 \Lambda _3^{\frac{3}{2}}}-\frac{\tilde{u}}{\Lambda _3^{\frac{1}{2}}}\right)  - \sum _{i=1}^3 \tilde{p} \, d\log(\tilde{p}+\tilde{m}_i ) \right\}.
\end{align}
$\tilde{m}_i$ $(i=1,\ldots, 3)$ are defined by
\begin{align}
\tilde{m}_1=&\frac{4 \tilde{M}^2}{3 \Lambda _3}+\tilde{c}_1, \qquad 
\tilde{m}_2= \frac{4 \tilde{M}^2}{3 \Lambda _3}+\tilde{c}_2, \qquad
\tilde{m}_3= \frac{4 \tilde{M}^2}{3 \Lambda _3}-(\tilde{c}_1+\tilde{c}_2) .
\end{align}
These parameters are interpreted as the mass parameters at the superconformal point.
We see that the scaling dimensions of $\tilde{u}$, $\tilde{M}$, $\tilde{C}_2$, $\tilde{C}_3$ are $\frac
{3}{2}$, $\frac{1}{2}$, $2$ and $3$, respectively.

We now study the SW periods for the AD theories associated with $SU(2)$ theory
with $N_f$ hypermultiplets.
We write  the SW curves in the form of
\begin{align}\label{eq:adc1}
\tilde{z}^2=\tilde{p}^3-\rho_{N_f} \tilde{p}-\sigma_{N_f}
\end{align}
for the $N_f$ AD theory.
Here $\rho_{N_f}$ and $\sigma_{N_f}$ are 
read off from (\ref{eq:swcadnf1}), (\ref{adc:nf2}) and (\ref{adc:nf3}).
We have normalized the
SW differential $\tilde{\lambda}_{SW}$ (\ref{addiff:nf1}), (\ref{eq:swdnf2}) and  (\ref{eq:swadnf3}) such that
\begin{align}
{\partial \over \partial\tilde{u}}\tilde{\lambda}_{SW}
={2d\tilde{p}\over \tilde{z}}.
\end{align}
The SW periods are defined by
\begin{align}
\tilde{\Pi}^{(0)}=(\tilde{a}^{(0)},\tilde{a}^{(0)}_D)=\left( \int_{\tilde{\alpha}} \tilde{\lambda}_{SW}, \int_{\tilde{\beta}} \tilde{\lambda}_{SW} \right),
\end{align}
where $\tilde{\alpha}$ and $\tilde{\beta}$ are canonical 1-cycles on the curve (\ref{eq:adc1}).
Differentiating the SW periods with respect to $\tilde{u}$, we have the period integral $\int {d\tilde{p}\over \tilde{z}}$ of the holomorphic differential $d\tilde{p}\over \tilde{z}$: 
\begin{align}
\omega &=\int_{\tilde{\alpha}} {d\tilde{p}\over \tilde{z}},\quad \omega_{D}=\int_{\tilde{\beta}} {d\tilde{p}\over \tilde{z}}.
\label{eq:period1}
\end{align}
As in the case of $SU(2)$ SQCD, 
the period integral is expressed in terms of  the hypergeometric functions of the argument:
\begin{align}
\tilde{w}_{N_f}&:=-{27\tilde{\Delta}_{N_f}\over 4\tilde{D}_{N_f}^3}=1-{27\sigma_{N_f}^2\over 4\rho_{N_f}^3}.
\end{align}
Here $\tilde{\Delta}_{N_f}$ and $\tilde{D}_{N_f}$ correspond to $\Delta$ in (\ref{discriminant:swc}) and $D$ in (\ref{eq:dd}) , respectively, which are defined by
\begin{align}
\tilde{\Delta}_{N_f}&=4\rho_{N_f}^3-27\sigma_{N_f}^2, \\
\tilde{D}_{N_f}&=-3\rho_{N_f}.
\end{align}

For example, we will evaluate the integrals (\ref{eq:period1}) around the point $\tilde{w}_{N_f}=0$, where  the $\tilde{\alpha}$-cycle is chosen as a vanishing cycle.
Using the quadratic and cubic transformation \cite{Erdelyi,Masuda:1996xj}, the periods 
are given by
\begin{align}
\omega^0(\tilde{w},\tilde{D}) =&2\pi \left(- \tilde{D}\right)^{-\frac{1}{4}} F\left( \frac{1}{12}, \frac{5}{12};1; \tilde{w} \right) ,\label{period_du:general2} \\
\omega^0 _D(\tilde{w},\tilde{D})=&-2i\pi \left(- \tilde{D}\right) ^{-\frac{1}{4}} \left(\frac{ 3\log 12}{ 2\pi } F\left( \frac{1}{12}, \frac{5}{12};1; \tilde{w} \right) -\frac{1}{2\pi } F_*\left( \frac{1}{12}, \frac{5}{12};1; \tilde{w} \right) \right), \label{periodD_du:general2}
\end{align}
where
$F(\alpha,\beta;\gamma;z)$ is the hypergeometric function.
$F_*(\alpha ,\beta ;1; z)$ is defined by
\begin{align} \label{Fstar:gamma1a}
F_*(\alpha ,\beta ; 1;z)= F( \alpha , \beta ;1;z) \log z+F_1( \alpha ,\beta ;1;z)
\end{align}
and 
\begin{align}
F_1(\alpha ,\beta ;1; z)=\sum _{n=0} ^\infty \frac{( \alpha )_n (\beta )_n}{(n!)^2} \sum _{r=0}^{n-1} \left( \frac{1}{\alpha +r} +\frac{1}{\beta +r}- \frac{2}{1+r} \right)z^n.
\end{align}
We have omitted the subscript $N_f$ of $\tilde{w}$ and $\tilde{D}$ for brevity.
Since the dual period has logarithmic divergence around $\tilde{w}=0$, it does not represent the
expansion around the superconformal point, where $\tilde{u}$ and $\tilde{M}$ have fractional scaling dimensions.

We will perform the analytic continuation of the solutions around $\tilde{w}=0$ to those of  $\tilde{w}=\infty$  by using the connection formula \cite{Erdelyi}
\begin{align} \label{transf:identity1}
\begin{split}
F(\alpha ,\beta ;\gamma ;z)=&\frac{ \Gamma (\gamma ) \Gamma (\beta -\alpha  ) }{\Gamma (\beta) \Gamma (\gamma -\alpha )} (1-z)^{-\alpha } F\left( \alpha ,\gamma -\beta ;\alpha -\beta +1;\frac{1}{1-z}\right) \\
&+\frac{ \Gamma (\gamma ) \Gamma (\alpha -\beta ) }{\Gamma (\alpha ) \Gamma (\gamma -\beta )}(1-z)^{-\beta } F\left( \beta ,\gamma -\alpha ;-\alpha +\beta +1;\frac{1}{1-z}\right),
\end{split}
\end{align}
where $|\arg (1-z)|< \pi$.
We then find that the periods (\ref{period_du:general2}) and (\ref{periodD_du:general2}) become
\begin{align} \label{period_du:general3}
\begin{split}
\omega^\infty(\tilde{w},\tilde{D}) =& 2\pi (-\tilde{D})^{-{1\over4}} \left( \frac{\Gamma \left(\frac{1}{3}\right)}{ \Gamma \left(\frac{5}{12}\right) \Gamma \left(\frac{11}{12}\right)} (1-\tilde{w})^{-{1\over12}}  F\left(\frac{1}{12},\frac{7}{12};\frac{2}{3};\frac{1}{1-\tilde{w}}\right) \right. \\
&\left. +\frac{\Gamma \left(-\frac{1}{3}\right)}{ \Gamma \left(\frac{1}{12}\right) \Gamma \left(\frac{7}{12}\right) } (1-\tilde{w})^{-{5\over 12}}  F\left(\frac{5}{12},\frac{11}{12};\frac{4}{3};\frac{1}{1-\tilde{w}}\right)  \right),
\end{split} 
\end{align}
\begin{align} \label{periodD_du:general3}
\begin{split}
\omega^\infty _D(\tilde{w},\tilde{D})=& 2i\pi (-\tilde{D})^{-{1\over4}} \left( \frac{(-1)^{\frac{5}{6}}\Gamma \left(\frac{1}{3}\right)}{ \Gamma \left(\frac{5}{12}\right) \Gamma \left(\frac{11}{12}\right)}  (1-\tilde{w}) ^{-\frac{1}{12}} F\left(\frac{1}{12},\frac{7}{12};\frac{2}{3};\frac{1}{1-\tilde{w}}\right) \right. \\
&+\frac{(-1)^{\frac{1}{6}}\Gamma \left(-\frac{1}{3}\right)}{ \Gamma \left(\frac{1}{12}\right) \Gamma \left(\frac{7}{12}\right) }\left. (1-\tilde{w})^{-\frac{5}{12}} F\left(\frac{5}{12},\frac{11}{12};\frac{4}{3};\frac{1}{1-\tilde{w}}\right) \right),
\end{split}
\end{align}
respectively.
Similarly we can perform the analytic continuation to the solutions around $\tilde{w}=1$.  By using
the connection formula
\begin{align}
\begin{split}
F \left( \alpha ,\beta ;\gamma ;z \right) =& \frac{(1-z)^{-\alpha -\beta +\gamma } \Gamma (\gamma ) \Gamma (\alpha +\beta -\gamma ) }{\Gamma (\alpha ) \Gamma (\beta )} F(\gamma -\alpha ,\gamma -\beta ;-\alpha -\beta +\gamma +1;1-z) \\
&+\frac{\Gamma (\gamma ) \Gamma (-\alpha -\beta +\gamma )}{\Gamma (\gamma -\alpha ) \Gamma (\gamma -\beta )}  F(\alpha ,\beta ;\alpha +\beta -\gamma +1;1-z),
\end{split}
\end{align}
we obtain expansion around $\tilde{w}=1$:
\begin{align}
\begin{split} \label{period_du:general4}
\omega^1 (\tilde{w},\tilde{D})=&\pi ^{-\frac{1}{2}} (-\tilde{D})^{-{1\over4}}\left( 6  \Gamma \left(\frac{5}{12}\right) \Gamma \left(\frac{13}{12}\right) F\left(\frac{1}{12},\frac{5}{12};\frac{1}{2};1-\tilde{w}\right) \right. \\
&\left. - (1-\tilde{w})^{1\over2} \Gamma \left(\frac{7}{12}\right) \Gamma \left(\frac{11}{12}\right) \, F\left(\frac{7}{12},\frac{11}{12};\frac{3}{2};1-\tilde{w}\right)  \right),
\end{split}\\
\begin{split} \label{periodD_du:general4}
\omega^1 _D(\tilde{w},\tilde{D})=&-i\pi ^{-\frac{1}{2}} (-\tilde{D})^{-{1\over4}}\left( 6  \Gamma \left(\frac{5}{12}\right) \Gamma \left(\frac{13}{12}\right) F\left(\frac{1}{12},\frac{5}{12};\frac{1}{2};1-\tilde{w}\right) \right. \\
&\left. + (1-\tilde{w})^{1\over2} \Gamma \left(\frac{7}{12}\right) \Gamma \left(\frac{11}{12}\right) \, F\left(\frac{7}{12},\frac{11}{12};\frac{3}{2};1-\tilde{w}\right)  \right).
\end{split} 
\end{align}

Based on these formulas, we discuss the SW periods for the AD theories.
For the $N_f=1$ theory, $\tilde{w}_1$ and $\tilde{D}_1$ are given by
\begin{align}
\tilde{w}_1&=1-{27 \tilde{u}^2\over 16\Lambda_1 \tilde{M}^3}, \label{wad:nf1} \\
\tilde{D}_1&=-3\Lambda_1 \tilde{M} .
\end{align}
The superconformal point corresponds to
$\tilde{w}'_1:={1\over 1-\tilde{w}_1}=0$.
Therefore eqs. (\ref{period_du:general3}) and (\ref{periodD_du:general3}) give the expansion 
around the superconformal point:
\begin{align}
{\partial \tilde{a}\over \partial \tilde{u}}=2\omega^\infty(\tilde{w}_1,\tilde{D}_1),
\quad 
{\partial \tilde{a}_D\over \partial \tilde{u}}=2\omega^\infty_D(\tilde{w}_1,\tilde{D}_1).
\end{align}
By integrating them over $\tilde{u}$,
we obtain the SW periods
\begin{align}
\tilde{a}^{(0)}=& \frac{3^{\frac{1}{2}} \Lambda _1^{\frac{3}{2}}}{2^{\frac{1}{2}} \cdot 5 \pi ^{\frac{1}{2}}}\left( \frac{\tilde{u}}{\Lambda _1^2} \right) ^{\frac{5}{6}}\left( 2^{\frac{8}{3}} \Gamma \left(\frac{1}{6}\right) \Gamma \left(\frac{1}{3}\right) F\left(-\frac{5}{12},\frac{1}{12};\frac{2}{3};\tilde{w}'_1\right)  \right. \nonumber \\
&\left. \qquad \qquad \qquad \qquad +15 \tilde{w}'_1{}^{\frac{1}{3}} \Gamma \left(-\frac{1}{6}\right) \Gamma \left(\frac{5}{3}\right) F\left(-\frac{1}{12},\frac{5}{12};\frac{4}{3};\tilde{w}'_1 \right)   \right) ,\label{ADperiod:nf1} 
\\
\tilde{a}_D ^{(0)} =& \frac{3^{\frac{1}{2}} \Lambda _1^{\frac{3}{2}}}{2^{\frac{1}{2}} \cdot 5 \pi ^{\frac{1}{2}}} \left( \frac{\tilde{u}}{\Lambda _1^2} \right)^{\frac{5}{6}} \left( -2^{\frac{8}{3}}(-1)^{\frac{1}{3}} \Gamma \left(\frac{1}{6}\right) \Gamma \left(\frac{1}{3}\right) F\left(-\frac{5}{12},\frac{1}{12};\frac{2}{3};\tilde{w}'_1\right)  \right. \nonumber \\
&\left. \qquad \qquad \qquad \qquad +15 (-1)^{\frac{2}{3}} \tilde{w}'_1{}^{\frac{1}{3}} \Gamma \left(-\frac{1}{6}\right) \Gamma \left(\frac{5}{3}\right) F\left(-\frac{1}{12},\frac{5}{12};\frac{4}{3};\tilde{w}'_1 \right)   \right). \label{ADperiodD:nf1}
\end{align} 
We note that the SW periods $\tilde{\Pi}^{(0)}$ satisfy
the Picard-Fuchs equation \cite{Ito:1999cc}
\begin{align} \label{PF:nf1}
\left(1-\tilde{w}'_1\right) \tilde{w}'_1 \frac{\partial ^2}{\partial \tilde{w}'_1{}^2 } \tilde{\Pi}^{(0)}+\frac{2}{3} \left(1-\tilde{w}'_1\right)  \frac{\partial }{\partial \tilde{w}'_1{} }\tilde{\Pi}^{(0)} +\frac{5}{144} \tilde{\Pi }^{(0)}=0.
\end{align}
From  (\ref{ADperiod:nf1}) and (\ref{ADperiodD:nf1}) we see that the SW periods scale as $\tilde{u}^{\frac{5}{6}}$.
Since the SW periods $a^{(0)}$ and $a^{(0)}_D$ have the scaling dimension one, the scaling dimension of $\tilde{u}$ and $\tilde{M}$
is  given by $\frac{6}{5}$ and $\frac{4}{5}$, respectively
 \cite{Argyres:1995xn}. 
The expansion  of the coupling constant $\tau ^{(0)}:=\frac{\partial _{\tilde{u}} \tilde{a}^{(0)}_D}{\partial _{\tilde{u}} \tilde{a} ^{(0)}}$ in $\tilde{w}_1'{}$ does not contain logarithmic terms, which implies that the theory is around the superconformal point.
The SW periods (\ref{ADperiod:nf1}) and (\ref{ADperiodD:nf1}) represent the expansions in the coupling $\tilde{M}$ with fixed $\tilde{u}$ in the scaling limit.
We note that the present expansions for $N_f$ theories are different from the results in the previous literatures \cite{Masuda:1996xj,Huang:2009md}, where  the coupling and the Casimir invariants are chosen to be zero, $\tilde{u}$ is small without taking the scaling limit. 
In \cite{Masuda:1996np} the expansion of the SW periods without taking the scaling limit has been presented.

For the $N_f=2$ theory, we have
\begin{align}
\tilde{w}_2&=1-{({27\over2} \tilde{C}_2\Lambda_2 -16 \tilde{M}^3+36 \tilde{M} \tilde{u})^2\over 432 \tilde{u}^3 },\\
\tilde{D}_2&=-3\tilde{u} .
\end{align}
The superconformal point corresponds to $\tilde{w}_2= 1$ or $\tilde{w}'_2:=1-\tilde{w}_2=0$.
Eqs. (\ref{period_du:general4}) and (\ref{periodD_du:general4}) provide the
expansion around the superconformal point:
\begin{align}
{\partial \tilde{a}\over \partial \tilde{u}}=2\omega^1(\tilde{w}_2,\tilde{D}_2),
\quad 
{\partial \tilde{a}_D\over \partial \tilde{u}}=2\omega^1_D(\tilde{w}_2,\tilde{D}_2).
\label{eq:nf2period}
\end{align}
Expanding them around $\tilde{w}'_2=0$, where $\frac{\tilde{M}^2}{\tilde{u} }\ll 1$ and $\frac{\tilde{C}_2\Lambda _2}{\tilde{u}^{\frac{3}{2}}} \ll 1$, and integrating over $\tilde{u}$, one obtains the SW periods, 
which are given by
\begin{align} \label{adperiod:nf2}
\tilde{a}^{(0)}=&\Lambda _2^{\frac{3}{2}}\left( \frac{\tilde{u}}{\Lambda _2^2} \right) ^{\frac{3}{4}} \left( \frac{2^4 \Gamma \left(\frac{5}{12}\right) \Gamma \left(\frac{13}{12}\right)}{3^{\frac{1}{4}} \pi ^{\frac{1}{2}}} 
-\frac{2^3\cdot 3^{\frac{1}{4}} \Gamma \left(\frac{7}{12}\right) \Gamma \left(\frac{11}{12}\right)}{ \pi ^{\frac{1}{2}} }\left( \frac{\tilde{M}^2}{\tilde{u}} \right)^{\frac{1}{2}} \right. \nonumber \\
&\left.
-\frac{3^{\frac{1}{2}} \Gamma \left(\frac{7}{12}\right) \Gamma \left(\frac{11}{12}\right)}{\pi ^{\frac{1}{2}}}\left( \frac{\tilde{C}_2^2\Lambda _2^2}{\tilde{u}^3} \right)^{\frac{1}{2}}
+\cdots   \right) ,
\end{align}
\begin{align} \label{adperiodD:nf2}
\tilde{a}_D^{(0)}=&\Lambda _2^{\frac{3}{2}}\left( \frac{\tilde{u}}{\Lambda _2^2} \right) ^{\frac{3}{4}} \left( -\frac{2^4i \Gamma \left(\frac{5}{12}\right) \Gamma \left(\frac{13}{12}\right)}{3^{\frac{1}{4}} \pi ^{\frac{1}{2}}} 
-\frac{2^3\cdot 3^{\frac{1}{4}}i \Gamma \left(\frac{7}{12}\right) \Gamma \left(\frac{11}{12}\right)}{ \pi ^{\frac{1}{2}} }\left( \frac{\tilde{M}^2}{\tilde{u}} \right)^{\frac{1}{2}} \right. \nonumber  \\
&\left.
-\frac{3^{\frac{1}{2}} i\Gamma \left(\frac{7}{12}\right) \Gamma \left(\frac{11}{12}\right)}{\pi ^{\frac{1}{2}}}\left( \frac{\tilde{C}_2^2\Lambda _2^2}{\tilde{u}^3} \right)^{\frac{1}{2}}
+\cdots   \right) .
\end{align}
We see again that the scaling dimensions of $\tilde{u}$, $\tilde{M}$ and $\tilde{C}_2$ are $\frac{4}{3} $, $\frac{2}{3}$ and $2$, respectively.
The expansions of the periods (\ref{adperiod:nf2}) and (\ref{adperiodD:nf2}) have no logarithmic behavior.

For the $N_f=3$ theory, we have
\begin{align}
\tilde{w}_3&=1-\frac{(-9\tilde{C}_3 \Lambda_3^3+18 \tilde{C}_2 \Lambda_3^2 \tilde{M}^2-16 \tilde{M}^6+72 \Lambda_3 \tilde{M}^3\tilde{u}+27 \Lambda_3^2 \tilde{u}^2)^2}{108\Lambda_3^6 \left( {\tilde{C}_2 \over2}+4 \frac{\tilde{M} \tilde{u}}{\Lambda _3} \right)^3},\\
\tilde{D}_3&=-3\left( {\tilde{C}_2\over2}+{4\tilde{M}\tilde{u}\over \Lambda_3} \right) .
\end{align}
The superconformal point corresponds to $\tilde{w}_3=\infty$ or $\tilde{w}'_3:={1\over 1-\tilde{w}_3}=0$.
Then (\ref{period_du:general3}) and (\ref{periodD_du:general3}) provides the periods around the superconformal point:
\begin{align}
{\partial \tilde{a}\over \partial \tilde{u}}=2\omega^\infty(\tilde{w}_3,\tilde{D}_3),
\quad 
{\partial \tilde{a}_D\over \partial \tilde{u}}=2\omega^\infty_D(\tilde{w}_3,\tilde{D}_3).
\label{eq:periodnf3}
\end{align}
Expanding these in $\tilde{w}'_3$, where 
$\frac{\tilde{M}^3}{\tilde{u}\Lambda _3} \ll 1$, $\frac{\tilde{C}_2^3 \Lambda _3^2}{\tilde{u}^4}\ll 1$ and $ \frac{\tilde{C}_3\Lambda _3 }{\tilde{u}^2}\ll 1$, 
and integrating (\ref{eq:periodnf3}) over $\tilde{u}$, we obtain the SW periods:
\begin{align}
\begin{split}
\tilde{a}^{(0)}=& \Lambda _3^{\frac{3}{2}}(-1)^{\frac{5}{6}}  \left( \frac{\tilde{u}}{\Lambda _3^2}\right) ^{\frac{2}{3}} \left( \frac{5 \Gamma \left(-\frac{5}{6}\right) \Gamma \left(\frac{1}{3}\right)}{2 \cdot 3^{\frac{1}{2}}\pi ^{\frac{1}{2}}} -\frac{2^3  \Gamma \left(-\frac{1}{3}\right) \Gamma \left(\frac{5}{6}\right)}{3^{\frac{1}{2}} \pi ^{\frac{1}{2}} }\left( \frac{\tilde{M}^3}{\tilde{u} \Lambda _3} \right) ^{\frac{1}{3}} 
\right.\\
&\left. + \frac{ \Gamma \left(-\frac{1}{3}\right) \Gamma \left(\frac{5}{6}\right)}{2 \cdot \pi ^{\frac{1}{2}} } \left( \frac{\tilde{C}_2^3 \Lambda _3^2}{\tilde{u}^4} \right)^{\frac{1}{3}} 
+\frac{ \Gamma \left(\frac{1}{6}\right) \Gamma \left(\frac{1}{3}\right)}{2^2 \cdot 3^{\frac{3}{2}}  \pi ^{\frac{1}{2}}} \left( \frac{\tilde{C}_3 \Lambda _3}{\tilde{u}^2} \right) 
+\cdots \right),
\end{split}
\end{align}
\begin{align}
\begin{split}
\tilde{a}_D^{(0)}=& \Lambda _3^{\frac{3}{2}}(-1)^{\frac{1}{6}} \left( \frac{\tilde{u}}{\Lambda _3^2}\right) ^{\frac{2}{3}} \left( \frac{5  \Gamma \left(-\frac{5}{6}\right) \Gamma \left(\frac{1}{3}\right)}{2 \cdot 3^{\frac{1}{2}}\pi ^{\frac{1}{2}}} -\frac{2^3 \Gamma \left(-\frac{1}{3}\right) \Gamma \left(\frac{5}{6}\right)}{3^{\frac{1}{2}} \pi ^{\frac{1}{2}} }\left( \frac{\tilde{M}^3}{\tilde{u} \Lambda _3} \right) ^{\frac{1}{3}} 
\right.\\
&\left.+ \frac{i \Gamma \left(-\frac{1}{3}\right) \Gamma \left(\frac{5}{6}\right)}{2 \cdot \pi ^{\frac{1}{2}} } \left( \frac{\tilde{C}_2^3 \Lambda _3^2}{\tilde{u}^4} \right)^{\frac{1}{3}} 
+\frac{\Gamma \left(\frac{1}{6}\right) \Gamma \left(\frac{1}{3}\right)}{2^2 \cdot 3^{\frac{3}{2}}  \pi ^{\frac{1}{2}}} \left( \frac{\tilde{C}_3 \Lambda _3}{\tilde{u}^2} \right) 
+\cdots \right).
\end{split}
\end{align}
It turns out that the scaling dimensions of $\tilde{u}$, $\tilde{M}$, $\tilde{C}_2$ and $\tilde{C}_3$ are $\frac{3}{2}$, $\frac{1}{2}$, $2$ and $3$, respectively.
As in the case of $N_f=1$ and $2$ theories, the expansion of the SW periods has no logarithmic term.

Although the SW curves for $N_f$ theories become a common cubic form,
their SW differentials  take  different forms 
due to the flavor symmetry. This means that we  need to introduce different quantization conditions for each $N_f$ as we will discuss in the next section.

\section{Quantum Seiberg-Witten curves and periods}
In this section 
we study the deformation of the SW periods in  the $\Omega$-background at the superconformal point for the $SU(2)$ gauge theory with $N_f(=1,2,3)$ hypermultiplets.
We take the the Nekrasov-Shatashvili (NS) limit such that one of the two deformation parameters $(\epsilon_1,\epsilon_2)$ of the $\Omega $ background is going to be zero.
The other parameter plays a role of the Planck constant $\hbar$. 
From the analysis of the $\Omega$-deformed low-energy effective action, the deformed periods in the NS limit are shown to satisfy the Bohr-Sommerfeld quantization condition \cite{Nekrasov:2009rc}:
\begin{align}
\oint \lambda _{\text{SW}} =i n \hbar ,\qquad (n \in \mathbf{Z}).
\end{align}
This condition also follows from the quantization of the SW curve, which is introduced by
the canonical quantization of the holomorphic symplectic structure defined by $d\lambda_{SW}$.
The quantum SW curve becomes the ordinary differential equation.
Its WKB solution gives the quantum correction to the SW periods,
which  can be represented in the form $\hat{\mathcal{O}}_k \Pi ^{(0)}$ for 
some differential operator   $\hat{\mathcal{O}}_k$  with respect to the moduli parameters.
In the following we will construct $\hat{\mathcal{O}}_2$ and  $\hat{\mathcal{ O}}_4$ explicitly 
and compute the second and fourth order corrections to the SW periods 
in $\hbar $ around the superconformal  point.

\subsection{$N_f=1$ theory}
We start with the $N_f=1$ theory.
The SW differential (\ref{addiff:nf1}) 
 defines a symplectic form $d\tilde{\lambda }_{\text{SW}}=d\tilde{z} \wedge d\tilde{p}$ on the $(\tilde{z}, \tilde{p})$ space.
We quantize the system by replacing the coordinate $\tilde{z} $ by the differential operator:
\begin{align} 
\tilde{z}=-i \hbar \frac{\partial }{\partial \tilde{p}}.
\end{align}
Then the SW curve becomes the Schr\"odinger type equation:
\begin{align} \label{qadc:nf1}
\left( - \hbar ^2 \frac{\partial ^2}{\partial \tilde{p}^2} +Q(\tilde{p}) \right) \Psi (\tilde{p}) =0,
\end{align}
where
\begin{align} \label{potential:nf1}
Q(\tilde{p}) = -\left( \tilde{p}^3 -\tilde{M} \Lambda _1 \tilde{p} -\frac{\Lambda _1}{2} \tilde{u}\right) .
\end{align}
We study the WKB solution to the equation (\ref{qadc:nf1}):
\begin{align} \label{wave_func}
\Psi (\tilde{p} )= \exp \left(\frac{i}{\hbar } \int ^{\tilde{p}} \Phi (y) dy\right),
\end{align}
where
\begin{align} \label{eq:wkb1}
\Phi (y) =\sum _{n=0}^\infty \hbar ^n \phi _n (y).
\end{align}
Substituting the expansion (\ref{eq:wkb1}) into (\ref{qadc:nf1}), one obtains the recursion relations 
for $\phi _n (\tilde{p})$'s. 
Note that $\phi _n(\tilde{p})$ for odd $n$ becomes a total derivative and  only $\phi _{n}(\tilde{p})$ 
for even $n$ contributes to the period integrals.
The first three $\phi _{2n}$'s are given by
\begin{align}
\phi_0(\tilde{p})&=i Q^{\frac{1}{2}},\label{recursion_relation} \\
\phi_2(\tilde{p} )&=\frac{i}{48}\frac{\partial_{\tilde{p}}^2 Q}{ Q^{\frac{3}{2}}}, \label{recursion_relation2}\\
\phi_4(\tilde{p} )&=-\frac{7i}{ 1536}\frac{(\partial_{\tilde{p}}^2 Q{})^2}{ Q^{\frac{7}{2}}}+\frac{i}{ 768}\frac{\partial_{\tilde{p}} ^4Q}{ Q^{\frac{5}{2}}}, \label{recursion_relation4}
\end{align}
up to total derivatives where $\partial _{\tilde{p}} := \frac{\partial}{\partial \tilde{p}}$.
We define the quantum SW periods
\begin{align}
\tilde{\Pi }=(\tilde{a}, \tilde{a}_D) =\left(\oint_{\tilde{\alpha}}  \Phi (\tilde{p}) d\tilde{p},\int_{\tilde{\beta} } \Phi(\tilde{p})d\tilde{p}\right) 
\end{align}
along the canonical 1-cycles $\tilde{\alpha }$ and $\tilde{\beta }$.
The periods are expanded in $\hbar $ as
\begin{align}
\tilde{\Pi }=\tilde{\Pi }^{(0)} +\hbar ^2 \tilde{\Pi }^{(2)} +\hbar ^4 \tilde{\Pi }^{(4)} +\cdots
\label{eq:qp1}
\end{align}
where $\tilde{\Pi} ^{(2n)} := \oint \phi _{2n } (\tilde{p}) d \tilde{p}$.
$\tilde{\Pi}^{(0)}$ is the classical SW period.
Similarly, we define $\tilde{a}^{(2n)}$ and $\tilde{a}_D^{(2n)}$ by 
\begin{align}
\tilde{a}=&\tilde{a}^{(0)}+\hbar ^2 \tilde{a}^{(2)} +\hbar ^4 \tilde{a}^{(4)} +\cdots ,\\
\tilde{a}_D=&\tilde{a}_D^{(0)}+\hbar ^2 \tilde{a}_D^{(2)} +\hbar ^4 \tilde{a}_D^{(4)} +\cdots .
\label{eq:qp2}
\end{align}
Substituting 
(\ref{potential:nf1}) into 
(\ref{recursion_relation2}) and (\ref{recursion_relation4}), one finds that 
\begin{align}
\phi _2(\tilde{p})
=&\frac{1}{\Lambda _1^2} \frac{\partial }{\partial \tilde{M}} \frac{\partial}{\partial \tilde{u}} \phi _0(\tilde{p}), \nonumber\\
\phi _4(\tilde{p})
=&\frac{7 }{10\Lambda _1^4}\frac{\partial ^2}{\partial \tilde{M}^2}\frac{\partial ^2}{\partial \tilde{u}^2} \phi _0 (\tilde{p})  . 
\label{eq:rec1}
\end{align}
The classical SW periods $\tilde{\Pi} ^{(0)}$
satisfy the Picard-Fuchs equation (\ref{PF:nf1}).
It is also found to 
satisfy the differential equation with respect to $\tilde{M}$ and $\tilde{u}$:
\begin{align} \label{massdiff:nf1}
\frac{\partial ^2}{\partial \tilde{M} \partial \tilde{u}} \tilde{\Pi } ^{(0)} =-\frac{3\tilde{u}}{2\tilde{M}} \frac{\partial ^2}{\partial \tilde{u}^2 } \tilde{\Pi }^{(0)} -\frac{1}{4\tilde{M}} \frac{\partial }{\partial \tilde{u}} \tilde{\Pi }^{(0)}.
\end{align}
From (\ref{eq:rec1}), 
the second and fourth order terms  satisfy
\begin{align}
\tilde{\Pi }^{(2)} =&\frac{1}{\Lambda _1^2}\frac{\partial }{\partial \tilde{M}} \frac{\partial}{\partial \tilde{u}} \tilde{\Pi }^{(0)}, \label{ADperiod2:nf1}\\
\tilde{\Pi }^{(4)} =&\frac{7}{10\Lambda _1^4} \frac{\partial ^2}{\partial \tilde{M}^2}\frac{\partial ^2}{\partial \tilde{u}^2} \tilde{\Pi }^{(0)} \label{ADperiod4:nf1}.
\end{align}

We note that the higher order corrections can be calculated by taking the scaling limit of those of the $N_f=1$ $SU(2)$ theory. 
The second and fourth order corrections  to the SW periods for the $N_f=1$ theory are given as \cite{Ito:2017iba}.
We can show that the formulas in \cite{Ito:2017iba} reduces to (\ref{ADperiod2:nf1}) and (\ref{ADperiod4:nf1}) in the scaling limit (\ref{rescaling:nf1}). 
The quantization conditions for the AD theories become different
although they take the same form for the SQCDs. Therefore it is  nontrivial to check that the scaling limit of the quantum SW periods of the SQCDs gives those of the AD theories.
In Section $4$, we will calculate the deformed SW periods around the superconformal point by using the relations (\ref{ADperiod2:nf1}) and (\ref{ADperiod4:nf1}) up to fourth order.

\subsection{$N_f=2$ theory}
Next we discuss the quantum  SW curve for the $N_f=2$ theory. 
We introduce a new variable $\xi$ by
\begin{align}
\tilde{p} =e^{ \xi} -\frac{2}{3}\tilde{M},
\end{align}
so that the SW differential (\ref{addiff:nf2}) becomes a canonical form
\begin{align}
\tilde{\lambda }_{\text{SW}} =  \tilde{z} d \xi.
\end{align}
The SW curve (\ref{adc:nf2}) takes  the form:
\begin{align}
\tilde{z}^2-\left( e^{3 \xi } -2 \tilde{M} e^{2 \xi } +e^{\xi } \left(\frac{4 \tilde{M}^2}{3}-\tilde{u}\right) -\frac{ \Lambda _2 \tilde{C}_2}{4} \right)=0.
\end{align}
Replacing  $\tilde{z}$ 
by the differential operator
\begin{align}
\tilde{z}=-i \hbar  \frac{\partial }{\partial \xi},
\end{align}
we obtain the quantum SW curve:
\begin{align}
\left( -\hbar ^2 \frac{\partial^2 }{\partial \xi ^2} +Q( \xi ) \right) \Psi (\xi)=0
\end{align}
where
\begin{align} \label{potential:nf2}
Q(\xi) =-\left( e^{3 \xi } -2 \tilde{M} e^{2 \xi } +e^{\xi } \left(\frac{4 \tilde{M}^2}{3}-\tilde{u}\right) -\frac{ \Lambda _2 \tilde{C}_2}{4} \right) .
\end{align}
We consider the WKB solution to 
the wave function $\Psi (\xi)$ which is defined by (\ref{wave_func}).
The leading term $\phi_0(\xi)$ 
in the expansion (\ref{eq:wkb1}) in $\hbar $ is given by $\phi _0(\xi) =\tilde{z}(\xi)$, 
which gives the classical 
 SW periods $\tilde{\Pi}^{(0)}=\int \phi _0(\xi) d\xi$.
One can show that
$(-\tilde{D}_2)^{\frac{1}{4}}\partial_{\tilde{u}}\tilde{\Pi}^{(0)}$ satisfies the Picard-Fuchs equation (\ref{eq:hyper1}).
$\tilde{\Pi}^{(0)}$ also satisfies 
the differential equation
\begin{align} \label{massdiff:nf2}
\frac{\partial ^2}{\partial \tilde{M} \partial \tilde{u}} \tilde{\Pi} ^{(0)} =L_2 \left( 4 \tilde{u} \frac{\partial ^2 }{\partial \tilde{u}^2} \tilde{\Pi} ^{(0)} + \frac{\partial  }{\partial \tilde{u}} \tilde{\Pi} ^{(0)} \right)
\end{align}
where
\begin{align}
L_2:= \frac{4 \left(4 \tilde{M}^2-3 \tilde{u}\right)}{27 \Lambda _2 \tilde{C}_2+24 \tilde{M} \tilde{u}-32 \tilde{M}^3} .
\end{align}
From (\ref{recursion_relation2}) and (\ref{recursion_relation4}), we find that the second and fourth order corrections are related to the classical SW period as
\begin{align}
\tilde{\Pi} ^{(2)}=&\left( \frac{1 }{4 } \frac{\partial }{\partial \tilde{M}}\frac{\partial }{\partial  \tilde{u} } +\frac{\tilde{M}}{3} \frac{\partial ^2}{\partial \tilde{u}^2}\right) \tilde{\Pi}^{(0)}, \label{ADperiod2:nf2}\\
\tilde{\Pi} ^{(4)}=& \left( \frac{7 \tilde{M}^2 }{90} \frac{\partial ^4}{\partial \tilde{u}^4}  +\frac{1}{20} \frac{\partial ^3}{\partial \tilde{u}^3} +\frac{7}{160 } \frac{\partial ^2}{\partial \tilde{u}^2}\frac{\partial ^2}{\partial \tilde{M}^2} +\frac{7 \tilde{M}}{60}\frac{\partial ^3}{\partial \tilde{u}^3} \frac{\partial }{\partial \tilde{M}} \right) \tilde{\Pi}^{(0)} \label{ADperiod4:nf2} .
\end{align}
Note that
(\ref{ADperiod2:nf2}) and (\ref{ADperiod4:nf2}) are defined up to the Picard-Fuchs equations. 
We also note that one can derive these relations from  those of $N_f=2$ $SU(2)$ theory, which are given by \cite{Ito:2017iba}.
We find that the second and fourth order formulas of the $N_f=2$ theory \cite{Ito:2017iba} lead to (\ref{ADperiod2:nf2}) and (\ref{ADperiod4:nf2}) after taking the scaling limit (\ref{rescaling:nf2}).

\subsection{$N_f=3$ theory}
Finally we study the quantum SW curve for the $N_f=3$ theory.
We introduce a new coordinate $\xi$ by
\begin{align}
\tilde{z}=-i \left( e^{\xi }+\frac{2 \tilde{M} \tilde{p}}{\Lambda _3^{\frac{1}{2}}}+\frac{4 \tilde{M}^3}{3 \Lambda _3^{\frac{3}{2}}}+\frac{\tilde{u}}{\Lambda _3^{\frac{1}{2}}} \right) ,
\end{align}
so that the SW differential (\ref{addiff:nf3}) becomes the 
canonical form
\begin{align}
\tilde{\lambda}_{SW}=i \Lambda _3 \left( \tilde{p}d\tilde{\xi}+\sum_{i=1}^3\tilde{p}d\log(\tilde{p}+\tilde{m}_i)\right).
\end{align}
Then 
the SW curve (\ref{adc:nf3}) 
can be written as
\begin{align}
&e^{2\xi} + (f_0 \tilde{p}+f_1) e^{\xi} +g(\tilde{p}) =0 ,\label{adc:nf3b} 
\end{align}
where
\begin{align}
\begin{split} \label{coefficient:nf3}
f_0=&\frac{4 \tilde{M}}{\Lambda _3^{\frac{1}{2}}}, \qquad
f_1= \frac{8 \tilde{M}^3}{3 \Lambda _3^{\frac{3}{2}}}+\frac{2 \tilde{u}}{\Lambda _3^{\frac{1}{2}}} ,\qquad
g(\tilde{p})= \tilde{p} ^3 -\rho _3 \tilde{p}-\sigma _3 +\left( \frac{2 \tilde{M} \tilde{p}}{\Lambda _3^{\frac{1}{2}}}+\frac{4 \tilde{M}^3}{3 \Lambda _3^{\frac{3}{2}}}+\frac{\tilde{u}}{\Lambda _3^{\frac{1}{2}}}  \right)^2 .
\end{split}
\end{align}
Replacing the coordinate $\xi $ by the differential operator 
\begin{align}
\xi =-i \hbar \frac{\partial }{\partial \tilde{p}},
\end{align}
one obtains the quantum SW curve. 
But we need to consider the ordering of the operators.
In general we can define the ordering of the operators by
\begin{align}
t \tilde{p} e^{-i\hbar \partial _{\tilde{p}}} \Psi ( \tilde{p} )+e^{-i \hbar \partial _{\tilde{p}}} \left( (1-t)\tilde{p} \Psi (\tilde{p})\right) =( \tilde{p}-i(1-t) \hbar )e^{-i \hbar \partial _{\tilde{p}}} \Psi (\tilde{p}),
\end{align}
parametrized by $t$  ($0\leq t\leq 1$).
We will use the $t={1\over2}$ prescription 
as in \cite{Zenkevich:2011zx}.
Then the quantum SW curve (\ref{adc:nf3b} ) takes the 
form
\begin{align} \label{qadc:nf3}
\left( \exp (-2i\hbar \partial _{\tilde{p}}) +\left( \frac{1}{2} f_0 \tilde{p} +f_1\right)\exp (-i \hbar \partial _{\tilde{p}}) +\exp (-i \hbar \partial _{\tilde{p}}) \frac{1}{2} f_0\tilde{p}+g(\tilde{p}) \right) \Psi (\tilde{p})=0.
\end{align}
We consider the WKB solution (\ref{wave_func}) to the quantum curve. 
The leading term is given by
$\phi _0({\tilde{p}}):=\xi (\tilde{p})$.
To discuss the higher order terms in $\hbar$, we rewrite the quantum curve by 
introducing
\begin{align} 
J(\alpha ):=&\exp \left( -\frac{i}{\hbar } \int ^{\tilde{p}} \Phi (y) dy \right) \exp \left( -i\hbar \alpha \partial _{\tilde{p}} \right) \exp \left( \frac{i}{\hbar } \int ^{\tilde{p}} \Phi (y)dy\right). \nonumber 
\end{align}
The quantum SW curve (\ref{qadc:nf3}) is written as
\begin{align} \label{qadc:nf3b}
J(2) +\left( f_0 \left( \tilde{p}-\frac{i}{2} \hbar \right)+f_1 \right) J(1) +g(x) =0.
\end{align}
Substituting (\ref{eq:wkb1}) into (\ref{qadc:nf3b}), 
we can determine $\phi_n(\tilde{p})$ in a recursive way.
$\phi_0(\tilde{p})$ is expressed as
\begin{align}
\phi_0(\tilde{p}) =\log \left( \frac{1}{2} \left( -f_0 \tilde{p}-f_1 + 2 \tilde{y}\right) \right)
\end{align}
which is equal to $\tilde{\xi}(\tilde{p})$.
Here $\tilde{y}$ is defined by
\begin{align}
\tilde{y}^2= \frac{1}{4} (f_0 \tilde{p} +f_1)^2-g(\tilde{p}).
\end{align}
$\phi_1(\tilde{p})$ is shown to be the total derivative:
\begin{align}
\phi _1( \tilde{p}) =\frac{\partial}{\partial \tilde{p}} \left( \frac{i}{2} \phi _0(\tilde{p}) +\frac{i}{4} \log 4\tilde{y} \right).
\end{align}
We can show that $\phi _3(\tilde{p})$ is also a total derivative.
$\phi_2$ and $\phi_4$ are found to be
\begin{align}
\phi_2(\tilde{p}) =& \frac{\left(-f_0 \tilde{p}-f_1\right) g''(\tilde{p})}{96 \tilde{y}^{3}}+\frac{f_0^2 \left(f_0 \tilde{p}+f_1\right)}{192 \tilde{y}^{3}}, \label{adc:p2:general} \\
\begin{split}
\phi _4(\tilde{p})=&g^{(4)}(\tilde{p}) \left(\frac{\left(f_0 \tilde{p}+f_1\right) g(\tilde{p})}{1536 \tilde{y}^{5}}+\frac{-f_0 \tilde{p}-f_1}{5760 \tilde{y}^{3}}\right)+g^{(3)}(\tilde{p}) \left(\frac{f_0 g(\tilde{p})}{480 \tilde{y}^{5}}+\frac{f_0}{720 \tilde{y}^{3}}\right) \\
&+g''(\tilde{p}) \left(-\frac{7 f_0^2 \left(f_0 \tilde{p}+f_1\right) g(\tilde{p})}{3072 \tilde{y}^{7}}-\frac{7 f_0^2 \left(f_0 \tilde{p}+f_1\right)}{7680 \tilde{y}^{5}}\right) \\
&+g''(\tilde{p})^2 \left(\frac{7 \left(f_0 \tilde{p}+f_1\right) g(\tilde{p})}{3072 \tilde{y}^{7}}+\frac{7 \left(f_0 \tilde{p}+f_1\right)}{7680 \tilde{y}^{5}}\right)
+\frac{7 f_0^4 \left(f_0 \tilde{p}+f_1\right) g(\tilde{p})}{12288 \tilde{y}^{7}}+\frac{7 f_0^4 \left(f_0 \tilde{p}+f_1\right)}{30720 \tilde{y}^{5}}, \label{adc:p4:general}
\end{split}
\end{align}
up to the total derivative.

For the classical SW periods $\tilde{\Pi}^{(0)}$, $(-\tilde{D}_3)^{\frac{1}{4}}\partial_{\tilde{u}}\tilde{\Pi}^{(0)}$
satisfies  the Picard-Fuchs equation (\ref{eq:hyper1}).
$\tilde{\Pi}^{(0)}$ also satisfies the differential equation with respect to $\tilde{M}$ and $\tilde{u}$:
\begin{align}
\frac{\partial ^2}{\partial \tilde{M} \partial \tilde{u} } \tilde{\Pi} ^{(0)}=b_3 \frac{\partial ^2}{\partial \tilde{u}^2} \tilde{\Pi} ^{(0)}+c_3 \frac{\partial }{\partial \tilde{u}} \tilde{\Pi} ^{(0)} 
\end{align}
where 
\begin{align}
b_3= &\frac{ 4 \tilde{M} \left(3 \Lambda _3 \tilde{M} \tilde{u}+4 \tilde{M}^4-3 \Lambda _3^2 \rho _3\right)\rho _3+27 \Lambda _3^2  \tilde{u} \sigma _3 }{3 \Lambda _3 \left(9 \Lambda _3 \tilde{M} \sigma _3 -4  \tilde{M}^3\rho _3 -3 \Lambda _3  \tilde{u}\rho _3 \right) }, \\
c_3=& \frac{\left(4 \tilde{M}^3+3 \Lambda _3 \tilde{u}\right){}^2-12 \Lambda _3^2 \tilde{M}^2 \rho _3 }{3 \Lambda _3 \left(9 \Lambda _3 \tilde{M} \sigma _3 -4  \tilde{M}^3\rho _3 -3 \Lambda _3  \tilde{u}\rho _3 \right) } .
\end{align}
$\rho _3$ and $\sigma _3$ are read off from (\ref{adc:nf3}).
Substituting (\ref{coefficient:nf3}) into (\ref{adc:p2:general}) and (\ref{adc:p4:general})
we find that formulas for  the second and fourth order corrections in $\hbar $:
\begin{align} 
\tilde{\Pi} ^{(2)} =&\left(-\frac{\tilde{M}^2}{12} \frac{\partial ^2}{\partial \tilde{u}^2} -\frac{\Lambda _3}{16}\frac{\partial }{\partial \tilde{u}}\frac{\partial }{\partial \tilde{M}}  \right) \tilde{\Pi} ^{(0)} ,\label{ADperiod2:nf3}
\\
\tilde{\Pi} ^{(4)}=&\left( \frac{7 \tilde{M}{}^4}{1440}\frac{\partial ^4}{\partial \tilde{u}{}^4} + \frac{ \Lambda _3 \tilde{M}}{192} \frac{\partial ^3}{\partial \tilde{u}{}^3} +\frac{7 \Lambda _3^2}{2560} \frac{\partial ^2}{\partial \tilde{u}{}^2} \frac{\partial ^2}{\partial \tilde{M}^2}   +\frac{7 \Lambda _3\tilde{M}{}^2}{960}  \frac{\partial ^3}{\partial \tilde{u}{}^3} \frac{\partial }{\partial \tilde{M}} \right) \tilde{\Pi}^{(0)}. \label{ADperiod4:nf3}
\end{align}
These formulas  
can be also obtained by taking scaling limit (\ref{rescaling:nf3}) of those in $N_f=3$ $SU(2)$ SQCD  \cite{Ito:2017iba}.

In the next section we will calculate the quantum corrections to the SW periods as an expansions in coupling constant and the mass parameters.

\section{Quantum SW periods around the superconformal point}

In the previous section we have constructed the quantum SW curves and the quantum SW periods of the AD theory,   which are obtained by acting the differential operators on the classical SW periods. 
In this section we will calculate an explicit form of the quantum SW periods around the superconformal point up to the fourth order in $\hbar$. 
We will consider the expansion in the coupling constant and the mass parameters of the AD theory.

\subsection{$N_f=1$ theory}
We first discuss the $N_f=1$ theory around the  superconformal point. 
Substituting (\ref{ADperiod:nf1}) and (\ref{ADperiodD:nf1}) into (\ref{ADperiod2:nf1})
 and changing the variables $(\tilde{u},\tilde{M})$ to ($\tilde{u}$,$\tilde{w}'_1$), 
the second order corrections to the SW periods are expressed in terms of hypergeometric function as
\begin{align}
\tilde{a}^{(2)}=&\frac{1}{2^{\frac{5}{2}}\cdot 3^{\frac{3}{2}} \pi ^{\frac{1}{2}} \Lambda _1^{\frac{7}{2}} }\left( \frac{\tilde{u}}{ \Lambda _1^2 }\right )^{-\frac{5}{6}} \left(  F^{(2)}_1(\tilde{w}'_1 ) -  F^{(2)}_2(\tilde{w}'_1  ) \right) ,\\
\tilde{a}_D^{(2)}=&\frac{1}{2^{\frac{5}{2}}\cdot 3^{\frac{3}{2}} \pi ^{\frac{1}{2}} \Lambda _1^{\frac{7}{2}} }\left( \frac{\tilde{u}}{ \Lambda _1^2 }\right )^{-\frac{5}{6}} \left( (-1)^{\frac{2}{3}} F^{(2)}_1(\tilde{w}'_1 ) + (-1)^{\frac{1}{3}} F^{(2)}_2(\tilde{w}'_1 ) \right),
\end{align}
where
\begin{align}
F^{(2)}_1(\tilde{w}'_1 )=& 2^{\frac{7}{3}}\cdot 3 \Gamma \left(\frac{2}{3}\right) \Gamma \left(\frac{5}{6}\right) \left(F\left(\frac{5}{12},\frac{11}{12};\frac{4}{3}; \tilde{w}'_1  \right)-5 F\left(\frac{11}{12},\frac{17}{12};\frac{4}{3};\tilde{w}'_1  \right) \right) ,\\
F^{(2)}_2(\tilde{w}'_1 ) =& -7  \tilde{w}'_1{}^{\frac{2}{3}} \Gamma \left(\frac{1}{6}\right) \Gamma \left(\frac{1}{3}\right) F\left(\frac{13}{12},\frac{19}{12};\frac{5}{3}; \tilde{w}'_1 \right).
\end{align}
Similarly, substituting (\ref{ADperiod:nf1}) and (\ref{ADperiodD:nf1}) into (\ref{ADperiod4:nf1})
 and changing the variables $(\tilde{u},\tilde{M})$ to ($\tilde{u}$,$\tilde{w}'_1$),
we  
find that the fourth order corrections to the SW periods (\ref{eq:qp2}) become
\begin{align}
\tilde{a}^{(4)}=& -\frac{7}{2^{\frac{43}{6}}\cdot 3^{\frac{5}{2}} \cdot 5 \pi ^{\frac{1}{2}} \Lambda _1^{\frac{17}{2}}  } \frac{\tilde{w}'_1{}^{\frac{1}{3}} }{\left( \tilde{w}'_1-1 \right) }\left( \frac{\tilde{u}}{ \Lambda _1^2 }\right )^{-\frac{5}{2}} \left( - F^{(4)}_1( \tilde{w}'_1 ) + F^{(4)} _2 (\tilde{w}'_1 ) \right)   ,\\
\tilde{a}_D^{(4)}=&-\frac{7}{2^{\frac{43}{6}}\cdot 3^{\frac{5}{2}} \cdot 5 \pi ^{\frac{1}{2}} \Lambda _1^{\frac{17}{2}}  } \frac{\tilde{w}'_1{}^{\frac{1}{3}} }{\left( \tilde{w}'_1-1 \right)  }\left( \frac{\tilde{u}}{ \Lambda _1^2 }\right )^{-\frac{5}{2}} \left( (-1)^{\frac{1}{3}} F^{(4)}_1( \tilde{w}'_1 ) +  (-1)^{\frac{2}{3}} F^{(4)} _2 (\tilde{w}'_1 ) \right) ,
\end{align}
where
\begin{align}
F^{(4)}_1(\tilde{w}'_1  )=& 2^3\cdot 7\cdot 13 \Gamma \left(\frac{1}{3}\right) \Gamma \left(\frac{7}{6}\right) \left((11 \tilde{w}'_1+13) F\left(\frac{19}{12},\frac{25}{12};\frac{5}{3};\tilde{w}'_1\right)-5 F\left(\frac{13}{12},\frac{19}{12};\frac{5}{3};\tilde{w}'_1\right)\right) ,\\
F^{(4)}_2(\tilde{w}'_1  ) =& 2^{\frac{1}{3}}\cdot 5 \cdot 11 \cdot 17\tilde{w}'_1{}^{\frac{1}{3}} \Gamma \left(\frac{2}{3}\right) \Gamma \left(\frac{5}{6}\right) \left((7 \tilde{w}'_1+17) F\left(\frac{23}{12},\frac{29}{12};\frac{7}{3};\tilde{w}'_1\right)-F\left(\frac{17}{12},\frac{23}{12};\frac{7}{3};\tilde{w}'_1\right)\right).
\end{align}
Expanding in $\tilde{w}'_1$ around $\tilde{w}'_1= 0$, the quantum SW periods become
\begin{align}
\tilde{a}=&  \Lambda _1^{\frac{3}{2}} \left( \frac{\tilde{u}}{\Lambda _1^2} \right)^{\frac{5}{6}} \left( -\frac{2^{\frac{7}{6}}   \Gamma \left(-\frac{5}{6}\right) \Gamma \left(\frac{1}{3}\right)}{3^{\frac{1}{2}} \pi ^{\frac{1}{2}}}  -\frac{7  \Gamma \left(-\frac{7}{6}\right) \Gamma \left(\frac{2}{3}\right)}{6^{\frac{1}{2}} \pi ^{\frac{1}{2}}}  \tilde{w}'_1{}^{\frac{1}{3}} +\cdots \right) \nonumber \\
&+ \frac{\hbar ^2}{\Lambda _1^{\frac{7}{2}} }\left( \frac{\tilde{u}}{\Lambda _1^{2}}\right)^{-\frac{5}{6}}  \left( -\frac{7  \Gamma \left(-\frac{7}{6}\right) \Gamma \left(\frac{2}{3}\right)}{2^{\frac{1}{6}} \cdot 3^{\frac{5}{2}} \pi ^{\frac{1}{2}} }+\cdots \right) \nonumber \\
&+ \frac{\hbar ^4}{\Lambda _1^{\frac{17}{2}} }\left( \frac{\tilde{u}}{\Lambda _1^{2}}\right)^{-\frac{5}{2}}\left(\frac{7^2\cdot 13 \Gamma \left(-\frac{5}{6}\right) \Gamma \left(\frac{1}{3}\right)}{2^{\frac{19}{6}}\cdot 3 ^{\frac{9}{2}} \pi ^{\frac{1}{2}} } \tilde{w}'_1{}^{\frac{1}{3}}  +\cdots  \right) +\cdots 
\label{aleading:nf1}, 
\end{align}
\begin{align}
\tilde{a}_D=& \Lambda _1^{\frac{3}{2}} \left( \frac{\tilde{u}}{\Lambda _1^2} \right)^{\frac{5}{6}} \left( \frac{2^{\frac{7}{6}}(-1)^{\frac{1}{3}}   \Gamma \left(-\frac{5}{6}\right) \Gamma \left(\frac{1}{3}\right)}{3^{\frac{1}{2}} \pi ^{\frac{1}{2}}}  -\frac{7 (-1)^{\frac{2}{3}}  \Gamma \left(-\frac{7}{6}\right) \Gamma \left(\frac{2}{3}\right)}{6^{\frac{1}{2}} \pi ^{\frac{1}{2}}} \tilde{w}'_1{}^{\frac{1}{3}} +\cdots \right) \nonumber \\
&+ \frac{\hbar ^2}{\Lambda _1^{\frac{7}{2}} }\left( \frac{\tilde{u}}{\Lambda _1^{2}}\right)^{-\frac{5}{6}}\left(- \frac{7 (-1)^{\frac{2}{3}} \Gamma \left(-\frac{7}{6}\right) \Gamma \left(\frac{2}{3}\right)}{2^{\frac{1}{6}} \cdot 3^{\frac{5}{2}} \pi ^{\frac{1}{2}} } +\cdots \right) \nonumber \\
&+\frac{\hbar ^4}{\Lambda _1^{\frac{17}{2}} }\left( \frac{\tilde{u}}{\Lambda _1^{2}}\right)^{-\frac{5}{2}} \left(-\frac{7^2\cdot 13 (-1)^{\frac{1}{3}}\Gamma \left(-\frac{5}{6}\right) \Gamma \left(\frac{1}{3}\right)}{2^{\frac{19}{6}}\cdot 3 ^{\frac{9}{2}} \pi ^{\frac{1}{2}} } \tilde{w}'_1{}^{\frac{1}{3}}  +\cdots  \right) +\cdots 
\label{aDleading:nf1}.
\end{align}
We define the effective coupling constant\footnote{Note that the present definition of the effective coupling constant is inverse of the one in \cite{Argyres:1995jj}.} $\tilde{\tau}$ of the deformed theory by
\begin{align} \label{eq:ec1}
\tilde{\tau} :=\frac{\partial _{\tilde{u}} \tilde{a}_D}{\partial _{\tilde{u}} \tilde{a}},
\end{align}
which is expanded in $\hbar$ as
\begin{align}
\tilde{\tau}&=\tilde{\tau}^{(0)}+\hbar^2 \tilde{\tau}^{(2)}+\hbar^4\tilde{\tau}^{(4)}+\cdots .
\end{align}
Substituting (\ref{aleading:nf1}) and (\ref{aDleading:nf1}) into (\ref{eq:ec1}) and 
expanding in $\hbar $, we find  
\begin{align}
\tilde{\tau} =&\left( -(-1)^{\frac{1}{3}} +\frac{3^{\frac{1}{2}}\cdot 7 i  \pi ^{\frac{1}{2}} \Gamma \left(-\frac{7}{6}\right)}{10 \Gamma \left(-\frac{5}{6}\right) \Gamma \left(\frac{1}{6}\right)} \tilde{w}'_1{}^{\frac{1}{3}}  + \cdots \right)\nonumber  \\ 
&+\frac{\hbar ^2}{\Lambda _1^5} \left( -\frac{2^{\frac{4}{3}}\cdot 3^{\frac{1}{2}} i \pi ^{\frac{1}{2}}  \Gamma \left(\frac{5}{6}\right)}{ \Gamma \left(\frac{1}{6}\right) \Gamma \left(-\frac{5}{6}\right)} \left( \frac{\tilde{u}}{\Lambda _1^2} \right) ^{-\frac{5}{3}} +\cdots \right) \nonumber \\
&+\frac{\hbar ^4}{\Lambda _1^{10}} \left( -\frac{2\cdot 3^{\frac{3}{2}} i \pi ^{\frac{1}{2}} \Gamma \left(\frac{5}{6}\right)^2 \Gamma \left(\frac{5}{3}\right)}{ \Gamma \left(-\frac{5}{6}\right)^2 \Gamma \left(\frac{1}{6}\right) \Gamma \left(\frac{1}{3}\right)}  \left(\frac{\tilde{u}}{\Lambda _1^2} \right) ^{-\frac{10}{3}} +\cdots \right)+\cdots .
\end{align}
We can express $\tilde{\tau}$ as a function of $\tilde{a}$ by solving (\ref{aleading:nf1}).
Then integrating it over $\tilde{a}$ twice, we obtain the free energy.
We find that the free energy at $\tilde{M}=0$ agrees with the one obtained from the E-string theory \cite{Sakai}.
We note that the present expansions for $N_f$ theories
in the coupling parameter are  different from  those in  the self-dual $\Omega$-background \cite{Huang:2009md}, where the expansions in 
the operator have been done with the zero coupling 
and without taking the scaling limit.

\subsection{$N_f=2$ theory}
We next compute the quantum corrections to the SW periods for the  $N_f=2$ theory.
From (\ref{ADperiod2:nf2}) and (\ref{eq:nf2period})
we find that the second order corrections are given by
\begin{align}
\tilde{a}^{(2)}= &-\frac{1}{2^{4}\cdot 3^{\frac{15}{4}} \pi ^{\frac{1}{2}} \Lambda _2^{\frac{3}{2}}} \left( \frac{\tilde{u}}{\Lambda _2^2}\right) ^{-\frac{3}{4}} \left( F_1 ^{(2)} ( \tilde{w}'_2 ) -F_2^{(2)} ( \tilde{w}'_2  ) \right),\\
\tilde{a}_D^{(2)}=&\frac{i}{2^{4}\cdot 3^{\frac{15}{4}} \pi ^{\frac{1}{2}} \Lambda _2^{\frac{3}{2}}} \left( \frac{\tilde{u}}{\Lambda _2 ^2}\right) ^{-\frac{3}{4}} \left( F_1 ^{(2)} (\tilde{w}'_2  ) +F_2^{(2)} (\tilde{w}'_2  ) \right),
\end{align}
where we have defined $\tilde{w}'_2=1-\tilde{w}_2$.
Here  $F_1^{(2)}(\tilde{w}'_2 )$ and $F_2^{(2)}(\tilde{w}'_2)$ are defined by
\begin{align}
F_1 ^{(2)} ( \tilde{w}'_2 )= & 3^2 \Gamma \left(\frac{1}{12}\right) \Gamma \left(\frac{5}{12}\right) \left(2^2\cdot 3^{\frac{1}{2}} \left( \frac{\tilde{M}^2}{\tilde{u}} \right)^{\frac{1}{2}} F\left(\frac{5}{12},\frac{13}{12};\frac{1}{2};\tilde{w}'_2\right)-5 \tilde{w}'_2{}^{\frac{1}{2}}  F\left(\frac{13}{12},\frac{17}{12};\frac{3}{2};\tilde{w}'_2 \right)\right),
\\
F_2 ^{(2)} ( \tilde{w}'_2 ) =& 6^2\Gamma \left(\frac{7}{12}\right) \Gamma \left(\frac{11}{12}\right) \left( 3  F\left(\frac{7}{12},\frac{11}{12};\frac{3}{2};\tilde{w}'_2 \right)+7   X^{(2)} F\left(\frac{11}{12},\frac{19}{12};\frac{3}{2};\tilde{w}'_2 \right) \right) ,
\end{align}
where
\begin{align}
X^{(2)}=-3+2\cdot 3^{\frac{1}{2}} \tilde{w}'_2{}^{\frac{1}{2}} \left( \frac{\tilde{M}^2}{\tilde{u}} \right)^{\frac{1}{2}}.
\end{align}
Expanding the second order terms in $\tilde{w}'_2$ around $\tilde{w}'_2=0$, where $\frac{\tilde{M}^2}{\tilde{u}}\ll 1$ and $\frac{\tilde{C}_2\Lambda_2}{\tilde{u}^{\frac{3}{2}}}\ll 1$, we obtain
\begin{align}
\tilde{a}^{(2)}=&  \frac{1 }{\Lambda _2^{\frac{3}{2}} } \left( \frac{\tilde{u}}{\Lambda _2^2} \right) ^{-\frac{3}{4}} \left( -\frac{3^{\frac{1}{4}} \Gamma \left(\frac{7}{12}\right) \Gamma \left(\frac{11}{12}\right)}{2 \pi ^{\frac{1}{2}}}
+\frac{\Gamma \left(\frac{1}{12}\right) \Gamma \left(\frac{5}{12}\right)}{2^4 \cdot 3^{\frac{5}{4}} \pi ^{\frac{1}{2}}}\left( \frac{\tilde{M}^2}{\tilde{u}} \right)^{\frac{1}{2}}
+\cdots  \right),\\
\tilde{a}_D^{(2)}=&\frac{1 }{\Lambda _2^{\frac{3}{2}} } \left( \frac{\tilde{u}}{\Lambda _2^2} \right) ^{-\frac{3}{4}} \left( -\frac{3^{\frac{1}{4}} i\Gamma \left(\frac{7}{12}\right) \Gamma \left(\frac{11}{12}\right)}{2 \pi ^{\frac{1}{2}}}
-\frac{i\Gamma \left(\frac{1}{12}\right) \Gamma \left(\frac{5}{12}\right)}{2^4 \cdot 3^{\frac{5}{4}} \pi ^{\frac{1}{2}}}\left( \frac{\tilde{M}^2}{\tilde{u}} \right)^{\frac{1}{2}}
+\cdots  \right) .
\end{align}

The fourth order corrections 
can be obtained in a similar manner. We find that
\begin{align}
\tilde{a}^{(4)}=& \frac{1}{2^{9}\cdot 3^{\frac{11}{4}} \cdot 5\pi ^{\frac{1}{2}} \Lambda _2^{\frac{9}{2}} } \frac{ 1 }{ \tilde{w}'_2{}^{\frac{1}{2}} (\tilde{w}'_2{} -1)^2} \left( \frac{\tilde{u}}{\Lambda _2^2}\right)^{-\frac{9}{4}} \left( F_1^{(4)} ( \tilde{w}'_2{} )-F_2^{(4)} (\tilde{w}'_2{} ) \right) ,\\
\tilde{a}_D^{(4)}=& -\frac{i}{2^{9}\cdot 3^{\frac{11}{4}} \cdot 5\pi ^{\frac{1}{2}} \Lambda _2^{\frac{9}{2}}}\frac{1 }{\tilde{w}'_2{}^{\frac{1}{2}}(\tilde{w}'_2{}-1 )^2}  \left( \frac{\tilde{u}}{\Lambda _2^2}\right)^{-\frac{9}{4}} \left( F_1^{(4)} (\tilde{w}'_2{} )+F_2^{(4)} (\tilde{w}'_2{} )\right),
\end{align}
where
\begin{align}
F_1^{(4)}(\tilde{w}'_2 ) =& \Gamma \left(\frac{1}{12}\right) \Gamma \left(\frac{5}{12}\right) \left( -14 X^{(4)}_1 F\left(\frac{1}{12},\frac{5}{12};\frac{1}{2};\tilde{w}'_2{}\right)+ X^{(4)}_2 F\left(\frac{5}{12},\frac{13}{12};\frac{1}{2};\tilde{w}'_2{}\right)\right) , \\
F_2^{(4)}(\tilde{w}'_2 ) =& 14\tilde{w}'_2{}^{\frac{1}{2}} \Gamma \left(\frac{7}{12}\right) \Gamma \left(\frac{11}{12}\right) \left( -2 X^{(4)}_1  F\left(\frac{7}{12},\frac{11}{12};\frac{3}{2};\tilde{w}'_2{}\right) +X^{(4)}_2 F\left(\frac{11}{12},\frac{19}{12};\frac{3}{2};\tilde{w}'_2{}\right)\right) .
\end{align}
Here the coefficients $X_1$ and $X_2$ are defined by
\begin{align}
\begin{split}
X^{(4)}_1  =& -2^2\cdot 3^{\frac{3}{2}}  \tilde{w}'_2{}^{\frac{1}{2}} \left(10 \tilde{w}'_2{} +11\right) +3 \left( \frac{\tilde{M}^2}{\tilde{u}} \right)^{\frac{1}{2}}  \left(377 \tilde{w}'_2{}+127\right)\\
&-2^3\cdot 3^{\frac{1}{2}} \left( \frac{\tilde{M}^2}{\tilde{u}} \right) \tilde{w}'_2{}^{\frac{1}{2}}  \left(13 \tilde{w}'_2{}+113\right) 
+28\left( \frac{\tilde{M}^2}{\tilde{u}} \right)^{\frac{3}{2}}  \left(13 \tilde{w}'_2{}+11\right),
\end{split}\\
\begin{split}
X^{(4)}_2 =&-3^{\frac{3}{2}} \tilde{w}'_2{}^{\frac{1}{2}}  \left(1345 \tilde{w}'_2{}+671\right)
+6 \left( \frac{\tilde{M}^2}{\tilde{u}} \right)^{\frac{1}{2}}  \left(520 \tilde{w}'_2{}^2+4639 \tilde{w}'_2{}+889\right)\\
& -2^2\cdot 3^{\frac{3}{2}} \left( \frac{\tilde{M}^2}{\tilde{u}} \right)  \tilde{w}'_2{}^{\frac{1}{2}} \left(593 \tilde{w}'_2{}+1423\right) 
+56  \left( \frac{\tilde{M}^2}{\tilde{u}} \right)^{\frac{3}{2}} \left(211 \tilde{w}'_2{}+77\right).  
\end{split}
\end{align}
Expanding the
fourth order corrections to the SW periods
in $\tilde{w}'_2$ around $\tilde{w}'_2=0$, where $\frac{\tilde{M}^2}{\tilde{u}}\ll 1$ and $\frac{\tilde{C}_2\Lambda_2}{\tilde{u}^{\frac{3}{2}}}\ll 1$, we get
\begin{align}
\tilde{a}^{(4)}=& \frac{1 }{\Lambda _2^{\frac{9}{2}} }\left( \frac{\tilde{u}}{\Lambda _2^2} \right) ^{-\frac{9}{4}} \left(-\frac{11 \Gamma \left(\frac{1}{12}\right) \Gamma \left(\frac{5}{12}\right)}{2^9\cdot 3^{\frac{5}{4}} \pi ^{\frac{1}{2}}}
-\frac{3^{\frac{1}{4}}\cdot  5\cdot 7 \Gamma \left(\frac{7}{12}\right) \Gamma \left(\frac{11}{12}\right)}{2^8 \pi ^{\frac{1}{2}}}\left( \frac{\tilde{M}^2}{\tilde{u}} \right)^{\frac{1}{2}} +\cdots \right),\\
\tilde{a}_D^{(4)}=&\frac{1}{\Lambda _2^{\frac{9}{2}} }\left( \frac{\tilde{u}}{\Lambda _2^2} \right) ^{-\frac{9}{4}} \left(\frac{11i \Gamma \left(\frac{1}{12}\right) \Gamma \left(\frac{5}{12}\right)}{2^9\cdot 3^{\frac{5}{4}} \pi ^{\frac{1}{2}}}
-\frac{3^{\frac{1}{4}}\cdot  5\cdot 7 i\Gamma \left(\frac{7}{12}\right) \Gamma \left(\frac{11}{12}\right)}{2^8 \pi ^{\frac{1}{2}}}\left( \frac{\tilde{M}^2}{\tilde{u}} \right)^{\frac{1}{2}}
+\cdots  \right).
\end{align}
The effective coupling constant $\tilde{\tau}$ is expanded in $\hbar$ as 
\begin{align} \label{ecc:nf2}
\begin{split}
\tilde{\tau} =&\left( -i -\frac{i \Gamma \left(\frac{7}{12}\right) \Gamma \left(\frac{11}{12}\right)}{ 3^{\frac{1}{2}} \Gamma \left(\frac{5}{12}\right) \Gamma \left(\frac{13}{12}\right)} \left( \frac{\tilde{M}^2}{\tilde{u}} \right)^{\frac{1}{2}} 
+\frac{i 3^{\frac{1}{2}}\Gamma \left(\frac{7}{12}\right) \Gamma \left(\frac{11}{12}\right)}{2^3 \Gamma \left(\frac{5}{12}\right) \Gamma \left(\frac{13}{12}\right)}\left( \frac{\tilde{C}_2^2\Lambda _2^2}{\tilde{u}^3} \right)^{\frac{1}{2}}
+\cdots  \right) \\
&+ \frac{\hbar ^2}{\Lambda _2^3}\left( \frac{\tilde{u}}{\Lambda _2^2} \right) ^{-\frac{3}{2}}  \left(\frac{3^{\frac{1}{2}}i  \Gamma \left(\frac{7}{12}\right) \Gamma \left(\frac{11}{12}\right)}{2^4 \Gamma \left(\frac{5}{12}\right) \Gamma \left(\frac{13}{12}\right)} 
+\frac{3^2 i \Gamma \left(\frac{7}{12}\right)^2 \Gamma \left(\frac{11}{12}\right)^2}{\Gamma \left(\frac{1}{12}\right)^2 \Gamma \left(\frac{5}{12}\right)^2}\left( \frac{\tilde{M}^2}{\tilde{u}} \right)^{\frac{1}{2}}
+\cdots  \right)  \\ 
&+\frac{\hbar ^4 }{\Lambda _2^6}\left( \frac{\tilde{u}}{\Lambda _2^2} \right) ^{-3}  \left( -\frac{3 i \Gamma \left(\frac{7}{12}\right)^2 \Gamma \left(\frac{11}{12}\right)^2}{2^9 \Gamma \left(\frac{5}{12}\right)^2 \Gamma \left(\frac{13}{12}\right)^2} \right. \\
&\left.
-\frac{3^{\frac{1}{2}} i \left(3 \Gamma \left(\frac{7}{12}\right)^3 \Gamma \left(\frac{11}{12}\right)^3+19 \pi ^2 \Gamma \left(\frac{5}{12}\right) \Gamma \left(\frac{13}{12}\right)\right)}{2^{10} \Gamma \left(\frac{5}{12}\right)^3 \Gamma \left(\frac{13}{12}\right)^3}\left( \frac{\tilde{M}^2}{\tilde{u}} \right)^{\frac{1}{2}}
+\cdots  \right)+\cdots .
\end{split}
\end{align}
It would be interesting to compare the free energy with that of the E-string theory, which is left for future work.

\subsection{$N_f=3$ theory}
We now discuss the $N_f=3$ case.
Using (\ref{ADperiod2:nf3}) and (\ref{eq:periodnf3}) 
we find that 
the second order corrections to the SW periods are given by
\begin{align}
\tilde{a}^{(2)}= &\frac{1}{2^{\frac{10}{3}} \cdot 3^{\frac{7}{2}} \pi ^{\frac{1}{2}} \tilde{w}'_3{}  \Lambda _3^{3}} \left( \frac{\tilde{u}}{\Lambda _3^2} \right)(- \sigma _3)^{-\frac{5}{6}} \left( 1+\frac{4}{3}\frac{\tilde{M}^3}{\tilde{u} \Lambda _3} \right) \left( F_1^{(2)}(\tilde{w}'_3{} )+ F_2^{(2)}(\tilde{w}'_3{} ) \right), \\
\tilde{a}_D^{(2)}=&\frac{i}{2^{\frac{10}{3}} \cdot 3^{\frac{7}{2}} \pi ^{\frac{1}{2}} \tilde{w}'_3{}   \Lambda _3^{3}} \left( \frac{\tilde{u}}{\Lambda _3^2} \right) (-\sigma _3)^{-\frac{5}{6}} \left( 1+\frac{4}{3}\frac{\tilde{M}^3}{\tilde{u} \Lambda _3} \right) \left( (-1)^{\frac{5}{6}} F_1^{(2)}(\tilde{w}'_3{} )+(-1)^{\frac{1}{6}} F_2^{(2)}(\tilde{w}'_3{} ) \right),
\end{align}
where 
$\tilde{w}'_3{} :=\frac{1}{1-\tilde{w}_3}$.
$F_1^{(2)}(\tilde{w}'_3{} )$ and $F_2^{(2)}(\tilde{w}'_3{} )$ are defined by
\begin{align}
F_1^{(2)}(\tilde{w}'_3{} )=&18 \Gamma \left(\frac{1}{6}\right) \Gamma \left(\frac{1}{3}\right) \left(  F\left(\frac{1}{12},\frac{7}{12};\frac{2}{3};\tilde{w}'_3{} \right) -X ^{(2)} F\left(\frac{7}{12},\frac{13}{12};\frac{2}{3};\tilde{w}'_3{} \right)  \right),\\
F_2^{(2)}(\tilde{w}'_3{} )=& -\frac{3\tilde{w}'_3{} }{2^{\frac{2}{3}}} \Gamma \left( -\frac{1}{6}\right) \Gamma \left( -\frac{1}{3}\right) \left(  F\left(\frac{5}{12},\frac{11}{12};\frac{4}{3};\tilde{w}'_3{} \right) -5 X ^{(2)} F\left(\frac{11}{12},\frac{17}{12};\frac{4}{3};\tilde{w}'_3{} \right)  \right).
\end{align}
Here $X^{(2)}$ is given by
\begin{align}
X ^{(2)} =1+ \frac{2^{\frac{2}{3}}\cdot 3\tilde{M} \Lambda _3}{(3\tilde{u} \Lambda _3+4 \tilde{M}^3)} (-\sigma _3)^{\frac{1}{3}} \tilde{w}'_3{}^{\frac{2}{3}}.
\end{align}
Expanding the second order corrections to the SW periods 
in $\tilde{w}'_3$, where
$\frac{\tilde{M}^3}{\tilde{u}\Lambda _3} \ll 1$, $\frac{\tilde{C}_2^3 \Lambda _3^2}{\tilde{u}^4}\ll 1$ and $ \frac{\tilde{C}_3\Lambda _3 }{\tilde{u}^2}\ll 1$, 
we obtain 
\begin{align}
\begin{split}
 \tilde{a}^{(2)}=& 
\frac{1}{\Lambda _3^{\frac{1}{2}}}  \left( \frac{\tilde{u}}{\Lambda _3^2}\right) ^{-\frac{2}{3}} \left(-\frac{(-1)^{\frac{1}{6}} \Gamma \left(\frac{2}{3}\right) \Gamma \left(\frac{5}{6}\right)}{2\cdot  3^{\frac{1}{2}} \pi^{\frac{1}{2}} }
+ \frac{\left(1+ 19(-1)^{\frac{1}{3}}\right) \Gamma \left(\frac{1}{6}\right) \Gamma \left(\frac{1}{3}\right)}{3^4 \pi^{\frac{1}{2}} } \left( \frac{\tilde{M}^3}{\tilde{u} \Lambda _3} \right) ^{\frac{2}{3}}    +\cdots\right) ,
\end{split}
\end{align}
\begin{align}
\begin{split}
\tilde{a}_D^{(2)}=&
\frac{1}{\Lambda _3^{\frac{1}{2}}}  \left( \frac{\tilde{u}}{\Lambda _3^2}\right) ^{-\frac{2}{3}} \left(-\frac{(-1)^{\frac{5}{6}} \Gamma \left(\frac{2}{3}\right) \Gamma \left(\frac{5}{6}\right)}{2\cdot  3^{\frac{1}{2}} \pi^{\frac{1}{2}} }
+ \frac{(-1)^{\frac{2}{3}}\left(1+19(-1)^{\frac{1}{3}}\right) \Gamma \left(\frac{1}{6}\right) \Gamma \left(\frac{1}{3}\right)}{3^4 \pi^{\frac{1}{2}} } \left( \frac{\tilde{M}^3}{\tilde{u} \Lambda _3} \right) ^{\frac{2}{3}} +\cdots\right) ,
\end{split}
\end{align}
The effective coupling constant 
is found to be
\begin{align}
\begin{split}
\tilde{\tau} =&\left( -(-1)^{\frac{1}{3}} -\frac{2^4 i \pi ^2}{\Gamma \left(\frac{1}{6}\right)^2 \Gamma \left(\frac{1}{3}\right)^2} \left( \frac{\tilde{M}^3}{\tilde{u} \Lambda _3} \right) ^{\frac{1}{3}}
-\frac{2 i \pi ^2}{\Gamma \left(\frac{1}{6}\right)^2 \Gamma \left(\frac{1}{3}\right)^2} \left( \frac{\tilde{C}_2^3 \Lambda _3^2}{\tilde{u}^4} \right)^{\frac{1}{3}} 
+\cdots \right) \\
& +\frac{\hbar ^2}{\Lambda _3^2 }
(-1)^{5\over6} \left( \frac{\tilde{u}}{\Lambda _3^2}\right) ^{-\frac{4}{3}}\left( -\frac{3^{\frac{1}{2}} \Gamma \left(\frac{2}{3}\right) \Gamma \left(\frac{5}{6}\right)}{5 \Gamma \left(-\frac{5}{6}\right) \Gamma \left(\frac{1}{3}\right)} 
-\frac{2^{\frac{10}{3}}\cdot 3^{\frac{1}{2}}  \Gamma \left(-\frac{1}{6}\right) \Gamma \left(\frac{5}{6}\right)^2}{5^2 \Gamma \left(-\frac{5}{6}\right)^2 \Gamma \left(\frac{1}{6}\right)}\left( \frac{\tilde{M}^3}{\tilde{u} \Lambda _3} \right) ^{\frac{1}{3}}   +\cdots  \right)  \\
&+\cdots .
\end{split}
\end{align}
We can calculate  the $\hbar^4$-order correction to the effective coupling constant in a similar way. 
The result is 
\begin{align}
\tilde{\tau}^{(4)}&=\frac{(-1)^{1\over6}}{\Lambda _3^4}
\left( \frac{\tilde{u}}{\Lambda _3^2}\right) ^{-\frac{8}{3}}\left( \frac{2^{\frac{4}{3}}\cdot 3^{\frac{1}{2}} \pi  \Gamma \left(\frac{5}{6}\right)^2}{5^2 \Gamma \left(-\frac{5}{6}\right)^2 \Gamma \left(\frac{1}{6}\right)^2}  
+\frac{2^3\cdot 3^{\frac{3}{2}}  \pi ^{\frac{1}{2}} \Gamma \left(-\frac{1}{6}\right) \Gamma \left(\frac{5}{6}\right)^3}{5^3 \Gamma \left(-\frac{5}{6}\right)^3 \Gamma \left(\frac{1}{6}\right)^2}\left( \frac{\tilde{M}^3}{\tilde{u} \Lambda _3} \right) ^{\frac{1}{3}}  + \cdots  \right).  
\end{align}

In summary, we have explicitly calculated the quantum corrections to the SW periods in terms of the hypergeometric functions up to the fourth orders in $\hbar$ for the AD theories of the $(A_1,A_2)$, $(A_1, A_3)$ and $(A_1,D_4)$-types.
\section{Conclusions and Discussions}
In this paper we studied the quantum SW periods around the superconformal point of ${\cal N}=2$ $SU(2)$ SQCD with $N_f=1,2,3$ hypermultiplets, which is deformed in the Nekrasov-Shatashvili limit of the $\Omega$-background. 
The  scaling limit around the superconformal point
gives the SW curves of the corresponding Argyres-Douglas theories.
The SW curves take the form of cubic
elliptic curve for  all $N_f$. 
But the SW differentials take the  different form, which introduce the different quantization condition. 
We have computed the quantum corrections to the SW periods up to the fourth order in $\hbar$, which are obtained from the classical periods by acting the differential operators with respect to the moduli
parameters. 
They are shown to agree with the scaling limit of the SW periods of the original SQCD.
We 
wrote down the explicit form of the quantum corrections in terms of hypergeometric functions.
It is interesting to explore the higher order corrections in $\hbar$. 
In particular the resurgence method  helps us to understand non-perturbative structure of the $\hbar$-corrections\cite{Basar:2015xna,Kashani-Poor:2015pca,Ashok:2016yxz, Basar:2017hpr}.
The SW curve for $N_f$ theory at the superconformal point are given by the scaling limit of the SW curve for the original SQCD. The  $SU(2)$ theory with the $N_f$ hypermultiplets are obtained by the decoupling limit for the $SU(2)$ theory with $N_f=4$ hypermultiplets. It is interesting to see the SW curve at the various superconformal points by combining both the scaling limit and the decoupling limit of the SW curve of the $SU(2)$ theory with $N_f=4$ hypermultiplets. 

So far we have studied the AD theories around the superconformal fixed point, where the SW periods
and the effective coupling constant are expanded in the Coulomb moduli parameter with fractional power. 
It would be interesting to study the $\hbar$-corrections to the beta functions around the conformal point \cite{Kubota:1997gb}.
Note that the moduli space of these AD theories contains the point, where one of the  periods
shows the logarithmic behavior around the point. 
It would be interesting  to describe the theory around the point by the Nekrasov partition function.

It is known that the four-dimensional theories in the NS limit are described by certain quantum integrable systems. The quantum corrections to the periods provide some data of the integrable systems.
For $N_f=1$ case, the curve describes the same AD theory as $SU(3)$ ${\cal N}=2$ super Yang-Mills theory \cite{Argyres:1995jj}, whose quantum curve is 
the Schr\"odinger equation with cubic polynomial potential.
In \cite{Ito:2017ypt}, using the ODE/IM correspondence (for a review see \cite{Dorey:2007zx}), it is shown that the the exponential of the quantum period can be regarded as the Y-function of the  quantum integrable model associated with the Yang-Lee edge singularity. 
It is interesting to study this relation further by computing further higher order corrections by using the ODE/IM correspondence.
It is also interesting to generalize  the quantum SW curve for the AD theories associated with higher rank gauge theories \cite{Masuda:1996np}.


\subsection*{Acknowledgements}
We would like to thank S. Kanno, H. Shu and K. Sakai for useful discussion.
The work of KI is supported
in part by Grant-in-Aid for Scientific Research 
15K05043, 18K03643 and  16F16735 from Japan Society for the Promotion of Science (JSPS).



\begin{thebibliography}{99}
 
 \bibitem{Argyres:1995jj}
  P.~C.~Argyres and M.~R.~Douglas,
  Nucl.\ Phys.\ B {\bf 448} (1995) 93
  [hep-th/9505062].

  
\bibitem{Argyres:1995xn} 
  P.~C.~Argyres, M.~R.~Plesser, N.~Seiberg and E.~Witten,
  Nucl.\ Phys.\ B {\bf 461}, 71 (1996)
  [hep-th/9511154].
  
  \bibitem{Eguchi:1996vu}
  T.~Eguchi, K.~Hori, K.~Ito and S.~K.~Yang,
  Nucl.\ Phys.\ B {\bf 471} (1996) 430
  [hep-th/9603002].  
  
\bibitem{Gaiotto:2009we} 
  D.~Gaiotto,
  JHEP {\bf 1208}, 034 (2012)
  [arXiv:0904.2715 [hep-th]].
  
\bibitem{Gaiotto:2009hg} 
  D.~Gaiotto, G.~W.~Moore and A.~Neitzke,
  arXiv:0907.3987 [hep-th].
  
\bibitem{Xie:2012hs} 
  D.~Xie,
  JHEP {\bf 1301}, 100 (2013)
  [arXiv:1204.2270 [hep-th]].
  
\bibitem{Beem:2013sza} 
  C.~Beem, M.~Lemos, P.~Liendo, W.~Peelaers, L.~Rastelli and B.~C.~van Rees,
  Commun.\ Math.\ Phys.\  {\bf 336}, no. 3, 1359 (2015)
  [arXiv:1312.5344 [hep-th]].
  
\bibitem{Liendo:2015ofa} 
  P.~Liendo, I.~Ramirez and J.~Seo,
  JHEP {\bf 1602}, 019 (2016)
  [arXiv:1509.00033 [hep-th]].

\bibitem{Cordova:2015nma} 
  C.~Cordova and S.~H.~Shao,
  JHEP {\bf 1601}, 040 (2016)
  [arXiv:1506.00265 [hep-th]].

\bibitem{Buican:2015ina} 
  M.~Buican and T.~Nishinaka,
  J.\ Phys.\ A {\bf 49}, no. 1, 015401 (2016)
  [arXiv:1505.05884 [hep-th]].

\bibitem{Nekrasov:2002qd}
  N.~A.~Nekrasov,
  Adv.\ Theor.\ Math.\ Phys.\  {\bf 7} (2004) 831
  [hep-th/0206161].

\bibitem{Nekrasov:2003rj}
  N.~Nekrasov and A.~Okounkov,
  hep-th/0306238.

\bibitem{Alday:2009aq}
  L.~F.~Alday, D.~Gaiotto and Y.~Tachikawa,
  Lett.\ Math.\ Phys.\  {\bf 91} (2010) 167
  [arXiv:0906.3219 [hep-th]].
  
\bibitem{Gaiotto:2009ma} 
  D.~Gaiotto,
  J.\ Phys.\ Conf.\ Ser.\  {\bf 462}, no. 1, 012014 (2013)
  [arXiv:0908.0307 [hep-th]].
  
  
  \bibitem{Huang:2009md}
  M.~x.~Huang and A.~Klemm,
  JHEP {\bf 1007} (2010) 083
  [arXiv:0902.1325 [hep-th]].  

\bibitem{Huang:2012kn} 
  M.~x.~Huang,
  JHEP {\bf 1206}, 152 (2012)
  [arXiv:1205.3652 [hep-th]].
  
  \bibitem{Bonelli:2016qwg}
  G.~Bonelli, O.~Lisovyy, K.~Maruyoshi, A.~Sciarappa and A.~Tanzini,
  arXiv:1612.06235 [hep-th].  
  
\bibitem{Huang:2011qx} 
  M.~x.~Huang, A.~K.~Kashani-Poor and A.~Klemm,
  Annales Henri Poincare {\bf 14}, 425 (2013)
  [arXiv:1109.5728 [hep-th]].
  
\bibitem{Sakai:2016jdi} 
  K.~Sakai,
  JHEP {\bf 1607}, 046 (2016)
  [arXiv:1603.09108 [hep-th]].
  
\bibitem{Nekrasov:2009rc}
  N.~A.~Nekrasov and S.~L.~Shatashvili,
  arXiv:0908.4052 [hep-th].  
  
\bibitem{Mironov:2009uv}
  A.~Mironov and A.~Morozov,
  JHEP {\bf 1004} (2010) 040
  [arXiv:0910.5670 [hep-th]].  
  
\bibitem{He:2010xa}
  W.~He and Y.~G.~Miao,
  Phys.\ Rev.\ D {\bf 82} (2010) 025020
  [arXiv:1006.1214 [hep-th]].  
  
     \bibitem{Zenkevich:2011zx}
  Y.~Zenkevich,
  Phys.\ Lett.\ B {\bf 701} (2011) 630
  [arXiv:1103.4843 [math-ph]].   
  
\bibitem{Ito:2017iba} 
  K.~Ito, S.~Kanno and T.~Okubo,
  JHEP {\bf 1708}, 065 (2017)
  [arXiv:1705.09120 [hep-th]].
  
\bibitem{Mironov:2009dv}
  A.~Mironov and A.~Morozov,
  J.\ Phys.\ A {\bf 43} (2010) 195401
  [arXiv:0911.2396 [hep-th]].
  
  \bibitem{Popolitov:2010bz}
  A.~Popolitov,
  Theor.  Math. Phys. {\bf 178} (2014) 239, 
  arXiv:1001.1407 [hep-th].  
  
\bibitem{Beccaria:2016wop} 
  M.~Beccaria,
  JHEP {\bf 1607}, 055 (2016)
  [arXiv:1605.00077 [hep-th]].
  
\bibitem{He:2016khf} 
  W.~He,
  arXiv:1608.05350 [math-ph].  
  
\bibitem{Hanany:1995na}
  A.~Hanany and Y.~Oz,
  Nucl.\ Phys.\ B {\bf 452} (1995) 283
  [hep-th/9505075].  
  
\bibitem{Brandhuber:1996ng}
  A.~Brandhuber and S.~Stieberger,
  Int.\ J.\ Mod.\ Phys.\ A {\bf 13} (1998) 1329
  [hep-th/9609130].
  
\bibitem{Ito:1995ga} 
  K.~Ito and S.~K.~Yang,
  Phys.\ Lett.\ B {\bf 366}, 165 (1996)
  [hep-th/9507144].
  
  \bibitem{Ohta:1996hq}
  Y.~Ohta,
  J.\ Math.\ Phys.\  {\bf 37} (1996) 6074
  [hep-th/9604051].
  
  \bibitem{Ohta:1996fr}
  Y.~Ohta,
  J.\ Math.\ Phys.\  {\bf 38} (1997) 682
  [hep-th/9604059].
 
   \bibitem{Erdelyi}
   A.~Erdelyi et al. ,
   "Higher Transcendental Functions", Vol. 1,
 MacGraw-Hill, New-York 

 
\bibitem{Masuda:1996xj} 
  T.~Masuda and H.~Suzuki,
  Int.\ J.\ Mod.\ Phys.\ A {\bf 12}, 3413 (1997)
  [Int.\ J.\ Mod.\ Phys.\ A {\bf 12}, 9700179 (1997)]
  [hep-th/9609066].
   
  
\bibitem{Ito:1999cc} 
  K.~Ito,
  Prog.\ Theor.\ Phys.\ Suppl.\  {\bf 135}, 94 (1999)
  [hep-th/9906023].

\bibitem{Masuda:1996np} 
  T.~Masuda and H.~Suzuki,
  Nucl.\ Phys.\ B {\bf 495}, 149 (1997)
  [hep-th/9612240]. 


\bibitem{Sakai} 
  K.~Sakai,
 private communication (March, 2018).  




\bibitem{Basar:2015xna} 
  G.~Basar and G.~V.~Dunne,
  JHEP {\bf 1502}, 160 (2015)
  [arXiv:1501.05671 [hep-th]].

\bibitem{Kashani-Poor:2015pca} 
  A.~K.~Kashani-Poor and J.~Troost,
  JHEP {\bf 1508}, 160 (2015)
  [arXiv:1504.08324 [hep-th]].
  
  \bibitem{Ashok:2016yxz}
  S.~K.~Ashok, D.~P.~Jatkar, R.~R.~John, M.~Raman and J.~Troost,
  JHEP {\bf 1607} (2016) 115
  [arXiv:1604.05520 [hep-th]].
  
\bibitem{Basar:2017hpr} 
  G.~Basar, G.~V.~Dunne and M.~Unsal,
  arXiv:1701.06572 [hep-th].

\bibitem{Kubota:1997gb}
  T.~Kubota and N.~Yokoi,
  Prog.\ Theor.\ Phys.\  {\bf 100} (1998) 423
  [hep-th/9712054].
  
 
  
\bibitem{Ito:2017ypt} 
  K.~Ito and H.~Shu,
  JHEP {\bf 1708}, 071 (2017)
  [arXiv:1707.03596 [hep-th]].
  


\bibitem{Dorey:2007zx}
  P.~Dorey, C.~Dunning and R.~Tateo,
  J.\ Phys.\ A {\bf 40} (2007) R205
  [hep-th/0703066].
  
 


  



  

  

  

  

 
 \end{thebibliography}
\end{document}